\documentclass{iopart}
\usepackage{epsfig}

\newcommand{\bth}{{\bar{\theta}}}
\newcommand{\bsi}{{\bar{\sigma}_+}}
\newcommand{\bqz}{{\bar{Q}_0}}
\newcommand{\bqp}{{\bar{Q}_+}}
\newcommand{\bc}{{\bar{C}_1}}
\newcommand{\bu}{{\bar{U}}}
\newcommand{\bw}{{\bar{W}}}
\newcommand{\by}{{\bar{Y}}}
\newcommand{\tqz}{{\tilde{Q}_0}}
\newcommand{\tqp}{{\tilde{Q}_+}}
\newcommand{\tc}{{\tilde{C}_1}}
\newcommand{\tct}{{\tilde{C}_2}}
\newcommand{\tu}{{\tilde{U}}}
\newcommand{\ty}{{\tilde{Y}}}
\newcommand{\sgn}[1]{{\,{\rm sgn}\left(#1\right)}}
\newcommand{\sqz}{{\epsilon}}
\newcommand{\omp}{{\Omega_\phi}}

\begin{document}

\title[Self-similar models with a fluid and a scalar field]{
  Self-similar spherically symmetric cosmological models with a
  perfect fluid and a scalar field} 
\author{Alan Coley\footnote{Department of Mathematics and Statistics,
    Dalhousie University, Halifax, Nova Scotia, B3H 3J5, Canada. E-mail:
    aac@mscs.dal.ca} \ and 
  Martin Goliath\footnote{Department of Physics, Stockholm University,
    Box 6730, S-113 85 Stockholm, Sweden. E-mail: goliath@physto.se}}

\begin{abstract}
  Self-similar, spherically symmetric cosmological models with a per\-fect
  fluid and a scalar field with an exponential potential are investigated.
  New variables are defined which lead to a compact state space, and
  dynamical systems methods are utilised to analyse the models. Due to
  the existence of monotone functions global dynamical results can be
  deduced. In particular, all of the future and past attractors for
  these models are obtained and the global results are discussed. The
  essential physical results are that initially expanding models always
  evolve away from a massless scalar field model with an initial
  singularity and, depending on the parameters of the models, either
  recollapse to a second singularity or expand forever towards a flat
  power-law inflationary model. The special cases in which there is no
  barotropic fluid and in which the scalar field is massless are
  considered in more detail in order to illustrate the asymptotic
  results. Some phase portraits are presented and the intermediate
  dynamics and hence the physical properties of the models are
  discussed. 
\end{abstract}

\pacs{0420, 0420J, 0440N, 9530S, 9880H}

\section{Introduction}

Scalar field cosmology is of importance in the study of the early
Universe and particularly in the investigation of inflation (during
which the universe undergoes a period of accelerated expansion
\cite{R1,Olive}). One particular class of inflationary cosmological
models are those with a scalar field and an exponential potential of
the form $V(\phi)=V_0 e^{-\kappa\phi}$, where $\kappa$ and $V_0$ are
constants. Models with an exponential scalar field potential arise
naturally in alternative theories of gravity, such as, for example,
scalar-tensor theories, and are currently of particular interest since
such theories occur as the low-energy limit in supergravity theories
\cite{T3,T5}.

A number of authors have studied scalar field cosmological models with
an exponential potential within general relativity. Homogeneous and 
isotropic Fried\-mann-Robertson-Walker (FRW) models were first studied by 
Halliwell \cite{R3} using phase-plane methods. Homogeneous but
anisotropic models of Bianchi types I and III (and Kantowski-Sachs
models) were studied by Burd and Barrow \cite{R5} in which they found
exact solutions and discussed their stability. Bianchi models of types
III and VI  were studied by Feinstein and Ib\'{a}\~{n}ez \cite{R8}, in which 
exact solutions were found. An analysis of Bianchi models, including
standard matter satisfying various energy conditions, was completed by
Kitada and Maeda \cite{R9,R92}. They found that the well-known power-law
inflationary solution is an attractor for all initially expanding
Bianchi models (except a subclass of the Bianchi type IX models which
will recollapse) when $\kappa^2<2$. 

The governing differential equations in spatially homogeneous Bianchi
cosmologies containing a scalar field with an exponential potential
reduce to a dynamical system when appropriate normalised variables are
defined; this dynamical system was studied in detail in
\cite{ColeyIbanezVanDenHoogen}. In a follow-up paper \cite{HCI} the
isotropisation of the Bianchi VII$_h$ cosmological models possessing a
scalar field with an exponential potential was further investigated;
in the case $\kappa^2>2$, it was shown that there is an open set of
initial conditions in the set of anisotropic Bianchi VII$_h$ initial
data such that the corresponding cosmological models isotropise
asymptotically. Hence, spatially homogeneous scalar field cosmological
models having an exponential potential with $\kappa^2>2$ can
isotropise to the future. The Bianchi type IX models have also been
studied in more detail \cite{HI}. 

Recently, cosmological models which contain both a perfect fluid and a
scalar field with an exponential potential have come under heavy
analysis \cite{bchio}.  One of the exact solutions found for these
models has the property that the energy density due to the scalar
field is proportional to the energy density of the perfect fluid,
hence these models have been labelled scaling cosmologies
\cite{CopelandLiddleWands}. These scaling solutions, which are
spatially flat isotropic models, are of particular physical interest.
For example, in these models a significant fraction of the current
energy density of the Universe may be contained in the scalar field
whose dynamical effects mimic cold dark matter. In
\cite{BillyardColeyVanDenHoogen} the stability of these cosmological
scaling solutions within the class of spatially homogeneous
cosmological models with a perfect fluid subject to the equation of
state $p=(\gamma-1)\mu$ (where $\gamma$ is a constant satisfying
$0<\gamma<2$) and a scalar field with an exponential potential was
studied. It was found that when $\gamma>2/3$, and particularly for
realistic matter with $\gamma\ge1$, the scaling solutions are
unstable; essentially they are unstable to curvature perturbations,
although they are stable to shear perturbations. Curvature scaling
solutions \cite{HCW} and anisotropic scaling solutions
\cite{ColeyIbanezOlasagasti} are also possible. In particular, in
\cite{HCW} homogeneous and isotropic spacetimes with non-zero spatial
curvature were studied. 

Clearly it is of interest to study more general cosmological models, and 
in this paper we shall comprehensively study the qualitative properties of 
the class of self-similar spherically symmetric models with a barotropic 
fluid and a non-interacting scalar field with an exponential potential.
Self-similar spherically symmetric perfect fluid models with a linear 
equation of state have been much studied in general relativity
\cite{Bogoyavlensky,CarrColey1998SS}. Carr \& Coley 
\cite{CarrColey1999classi} have presented a complete asymptotic 
classification of such solutions and, by reformulating the field
equations for these models, Goliath \emph{et al} \cite{GNU1,GNU2} have
obtained a compact three-dimensional state space representation of the
solutions which leads to another complete picture of the solution
space. Recently, these models have been further studied in a combined
approach \cite{ccgnu}. The present analysis can be thought of as a
natural extension of this recent work \cite{GNU1,GNU2,ccgnu}. 

The Kantowski-Sachs models appear as a limiting case of the
class of spherically symmetric models under investigation, and hence
this analysis complements recent analyses of spatially homogeneous
Bianchi models \cite{bchio}. Models with {\it positive spatial curvature}
have attracted less attention than Bianchi models with
{\it zero spatial curvature} or {\it negative spatial curvature} since
they are more complicated mathematically. However, the properties of
positive-curvature FRW models \cite{R3,HCW,Aber} and Kantowski-Sachs
models \cite{R5} have been studied previously. 

In the next section we shall describe the governing equations of the
class of models under investigation. In section \ref{sec:sss}, compact
variables are defined and the resulting dynamical system is derived in
the case of spatial self-similarity. A monotone function is
obtained. The equilibrium points and their local stability is
discussed in section \ref{sec:ssseq}. The timelike self-similar case
is then considered in sections \ref{sec:tss} and \ref{sec:tsseq}. The
special equilibrium points for values of ``extreme tilt'' are
discussed separately in section \ref{sec:extreme}. The global results
and a discussion is given in section \ref{sec:global}. Applications in
the absence of a barotropic fluid and in the further subcase of a
massless scalar field are discussed, respectively, in sections
\ref{sec:fluidvac} and \ref{sec:massless}, partially to illustrate the
early-time and late-time behaviour of the models.

\section{Governing equations}

We shall consider spherically symmetric similarity solutions in which
the source for the gravitational field is a perfect fluid and a 
non-interacting scalar field with an exponential potential in which the 
total energy-momentum tensor is given by:
\begin{equation}
  T_{ab}=(T_{{\rm pf}\,ab}+T_{{\rm sf}\,ab}) , 
\end{equation}
where 
\begin{equation}
  T_{{\rm pf}\,ab}=\mu u_au_b+p\left(u_au_b+g_{ab}\right) . 
\end{equation}
The perfect fluid obeys the equation of state
\begin{equation}\label{eq:eos}
  p=(\gamma-1)\mu ,
\end{equation} 
where $\gamma$ is a constant satisfying $1<\gamma<2$. 
The scalar-field contribution is given by
\begin{equation}
  T_{{\rm sf}\,ab}=\phi_{,a}\phi_{,b}-
  \left[\frac{1}{2}\phi_{,c}\phi^{,c} + V(\phi)\right]g_{ab} , 
\end{equation}
where
\begin{equation}
  V(\phi)=V_0e^{-\kappa\phi} .
\end{equation}

Since the fluid and scalar field are non-interacting, we have the
following separate conservation laws: 
\begin{equation}\label{eq:conserv}
  \nabla_aT_{\rm pf}^{ab}=0=\nabla_aT_{\rm sf}^{ab} .
\end{equation}
The spacetime is self-similar and consequently admits a homothetic
vector $\eta^a$. This implies that the matter fields must be of a
particular form. Thus, a barotropic fluid must have an equation of
state of the form (\ref{eq:eos}) \cite{CahillTaub}. The 
energy-momentum tensor of the scalar field must satisfy:
\begin{equation}
  {\cal L}_\eta T_{{\rm sf}\,ab}=0.
\end{equation} 
This implies that
\begin{eqnarray}  
  \phi&=&\Phi(\xi)+\frac{2}{\kappa}\eta , \label{eq:phi} \\
  V&=&e^{-2\eta}{\cal V}(\Phi), \\
  {\cal V}(\Phi)&=&V_0e^{-\kappa\Phi(\xi)} ,    
\end{eqnarray}
where $\eta$ is the variable defined by the homothetic vector $\eta^a$,
and $\xi$ is the similarity variable. When the similarity variable is 
timelike, we will use the notation $t\equiv\xi,x\equiv\eta$. 
The homothetic vector then is spacelike, and we denote this as the
{\em spatially self-similar} case. When the homothetic vector is timelike, we 
have the {\em timelike self-similar} case, for which $t\equiv\eta,x\equiv\xi$.
A dot denotes differentiation with respect to the similarity variable.
Finally, we shall define the new variable:
\begin{equation}
  X=\frac{1}{\sqrt{2}}\dot{\Phi} ,
\end{equation}
and for convenience we introduce the new constant
\begin{equation}
  k\equiv\frac{\sqrt{2}}{\kappa} .
\end{equation}

\section{Spatially self-similar case}\label{sec:sss}

In the spatially self-similar case, the line element can be written
\cite{Bogoyavlensky} 
\begin{eqnarray}
  ds^2&=&e^{2x}d\bar{s}^2 , \\
  d\bar{s}^2&=&-dt^2+D_1^2dx^2+D_2^2d\Omega^2 , \\
  D_1&=&e^{\beta^0-2\beta^+} , \quad 
  D_2=e^{\beta^0+\beta^+} .
\end{eqnarray}
The kinematic quantities of the congruence normal to the symmetry
surfaces are related to the Misner variables ($\beta^0$, $\beta^+$) as
follows: 
\begin{equation}
  \theta=3\dot{\beta}^0 , \quad 
  \sigma_+=3\dot{\beta}^+ .
\end{equation}
Following \cite{GNU1}, we will work with boosted kinematic quantities 
($\bth,\bsi$) 
\begin{equation}
  \theta=\frac{1}{\sqrt{3}}\left(2\bth+\bsi\right) , \quad
  \sigma_+=\frac{1}{\sqrt{3}}\left(\bth+2\bsi\right) .
\end{equation}
The reason for this is that it simplifies the constraint obtained from the 
non-vanishing off-diagonal component of the field equations.
The metric functions $B_1\equiv D_1^{-1}$ and $B_2\equiv D_2^{-1}$ 
then have the following evolution equations:
\begin{eqnarray}
  \dot{B_1}&=&\frac{1}{\sqrt{3}}\bsi B_1 , \\
  \dot{B_2}&=&-\frac{1}{\sqrt{3}}(\bth+\bsi)B_2 .
\end{eqnarray}

The physical quantities associated with the perfect fluid are as follows:
\begin{eqnarray}
  u_{\rm pf}^a&=&\frac{e^{-x}}{\sqrt{1-v^2}}
  \left(1,ve^{2\beta^+-\beta^0},0,0\right) ,\\
  \mu&=&\frac{1-v^2}{1+(\gamma-1)v^2}e^{-2x}\mu_n ,
\end{eqnarray}
where $v$ is the tilt variable, and $\mu_n$ is the energy density
along the normal congruence. From the conservation equations
(\ref{eq:conserv}) for the perfect fluid, we have: 
\begin{eqnarray}
  \dot{\mu}_n&=&-\frac{\gamma}{\sqrt{3}\left[1+(\gamma-1)v^2\right]}
  \left[2\bth+(1-v^2)\bsi+2\sqrt{3}vB_1\right]\mu_n , \\
  \dot{v}&=&\frac{1-v^2}{\sqrt{3}\gamma\left[1-(\gamma-1)v^2\right]}
  \left\{\gamma\left[2(\gamma-1)\bth+\gamma\bsi\right]v
  \right. \nonumber \\ && \left.
  +\sqrt{3}\left[(\gamma-1)(3\gamma-2)v^2-(2-\gamma)\right]B_1\right\} ,
\end{eqnarray}
and the conservation equation for the scalar field yields
\begin{eqnarray}
  \dot{X}&=&-\frac{1}{\sqrt{3}}(2\bth+\bsi)X
  +2kB_1^2+\frac{{\cal V}}{k} .
\end{eqnarray}
The Einstein field equations then yield the following:\\
{\em The Friedmann equation}:
\begin{eqnarray}
  \mu_n&=&\frac{1}{3}\left[\bth^2-\bsi^2-3(1+k^2)B_1^2
  +3B_2^2-3X^2-3{\cal V}\right] \label{eq:friedmannSSS} .
\end{eqnarray}
{\em Constraint equation}:
\begin{eqnarray}
  0&=&\gamma v\mu_n-\frac{2}{\sqrt{3}}\left[1+(\gamma-1)v^2\right]
  \left(\bsi+\sqrt{3}kX\right)B_1 .
\end{eqnarray}
{\em Evolution equations for $\bth$ and $\bsi$}:
\begin{eqnarray}
  \dot{\bth}&=&-\frac{1}{\sqrt{3}}\left[\bth^2+\bsi^2+\bth\bsi
  -3(1+k^2)B_1^2+3X^2-3{\cal V} \right. \nonumber \\
  &+&\left.\frac{3(\gamma-1)(1-v^2)}{1+(\gamma-1)v^2}\mu_n\right] , \\
  \dot{\bsi}&=&-\frac{1}{\sqrt{3}}\left[\bsi^2+2\bth\bsi
  +6k^2B_1^2+3{\cal V}
  \right. \nonumber \\ && \left.
  +\frac{3}{2}\frac{(2-\gamma)+(3\gamma-2)v^2}{1+(\gamma-1)v^2}\mu_n \right] .
\end{eqnarray}

From equation (\ref{eq:friedmannSSS}), by demanding $\mu_n\geq0$, a
dominant quantity 
\begin{eqnarray}
  \by&=&\sqrt{\bth^2+3B_2^2} 
\end{eqnarray}
is identified. Thus, we define bounded variables according to
\begin{eqnarray}
  &&\bqz=\frac{\bth}{\by} , \quad
  \bqp=\frac{\bsi}{\by} , \quad
  \bc=\frac{\sqrt{3}B_1}{\by} , \quad 
  \bu=\frac{\sqrt{3}X}{\by} , \quad
  \bw=\frac{\sqrt{3{\cal V}}}{\by} .
\end{eqnarray}
Defining an appropriate density parameter with respect to $\mu_n$, 
the Friedmann equation takes the form:
\begin{eqnarray}
  \Omega_n&=&\frac{3\mu_n}{\by^2} \\
  &=&1-\bqp^2-(1+k^2)\bc^2-\bu^2-\bw^2 , 
\end{eqnarray}
while the constraint equation becomes
\begin{eqnarray}\label{eq:sss-constr}
  0&=&\gamma v\Omega_n-2\left[1+(\gamma-1)v^2\right](\bqp+k\bu)\bc .
\end{eqnarray}
By defining a new independent variable
\begin{eqnarray}
  '&=&\frac{d}{d\tau}=\frac{\sqrt{3}}{\by}\frac{d}{dt} ,
\end{eqnarray}
the evolution equation for $\by$ 
\begin{eqnarray}
  \by'&=&-\left\{\bqp+\bqz\left[2(\bqp^2+\bu^2)+
  \frac{\gamma}{1+(\gamma-1)v^2}\Omega_n\right]\right\}\by 
\end{eqnarray}
decouples, and we are left with a reduced set of evolution equations:
\begin{eqnarray}
  \bqz'&=&-(1-\bqz^2)\left[1-2(1-\bqp^2-\bu^2)+
  \frac{\gamma}{1+(\gamma-1)v^2}\Omega_n\right] , \\
  \bqp'&=&\bqz\bqp\left[-2(1-\bqp^2-\bu^2)+
  \frac{\gamma}{1+(\gamma-1)v^2}\Omega_n\right]-\nonumber\\
  &&\left[2k^2\bc^2+\bw^2+\frac{1}{2}\frac{(2-\gamma)+(3\gamma-2)v^2} 
  {1+(\gamma-1)v^2}\Omega_n\right] , \\
  \bc'&=&2\bc\left[\bqp+\bqz(\bqp^2+\bu^2)+\frac{1}{2}
  \frac{\gamma}{1+(\gamma-1)v^2}\bqz\Omega_n\right] , \label{eq:bc} \\
  v'&=&\frac{1-v^2}{\gamma\left[1-(\gamma-1)v^2\right]}
  \left\{\gamma\left[2(\gamma-1)\bqz+\gamma\bqp\right]v+
  \right. \nonumber \\ && \left.
  +\left[(\gamma-1)(3\gamma-2)v^2-(2-\gamma)\right]\bc\right\} , \\
  \bu'&=&\bqz \bu\left[-2(1-\bqp^2-\bu^2)+\frac{\gamma}{1+(\gamma-1)v^2}
  \Omega_n \right]
  \nonumber \\ &&
  +2k\bc^2+\frac{\bw^2}{k} , \\
  \bw'&=&\frac{\bw}{k}\left\{k\bqp-\bu+k\bqz
  \left[2(\bqp^2+\bu^2)+\frac{\gamma}{1+(\gamma-1)v^2}\Omega_n\right]\right\}
    \label{eq:bw} .
\end{eqnarray}
This system is invariant under the transformation
\begin{equation}
  (\tau,\bqz,\bqp,\bc,v,\bu,\bw)\rightarrow
  (-\tau,-\bqz,-\bqp,\bc,-v,-\bu,\bw) .
\end{equation}
Furthermore, $\bw\geq0$, and by noting the invariance under the
transformation $(\bc,v)\rightarrow(-\bc,-v)$, we can without loss of 
generality restrict the analysis to $\bc\geq0$.
There is also an auxiliary evolution equation for $\Omega_n$:
\begin{eqnarray}
  \Omega_n'&=&-\Omega_n\left\{\frac{\gamma}{1+(\gamma-1)v^2}
  \left[2\bqz+(1-v^2)\bqp+2v\bc\right]+2\frac{\by'}{\by}\right\} .
\end{eqnarray}
In appendix \ref{app:q}, expressions for some important fluid
quantities are given.

\subsection{Invariant submanifolds}

A number of invariant submanifolds can be identified:
\begin{itemize}
  \item Plane symmetric: $\bqz=\pm1$, which implies $B_2=0$ (see \cite{GNU1}).
  \item  Non-self-similar Kantowski-Sachs: $\bc=0,v=0$  
  (see \cite{GNU1}).
  \item Massless scalar field: $\bw=0$.
  \item No perfect fluid: $\Omega_n=0$ (and $v$ decouples).
  \item No scalar field: $\bu=0$, $\bw=0$ and $k=0$.
\end{itemize}

The case with no scalar field was studied in \cite{GNU1}, where the
global dynamics of these models was investigated in detail. The
resulting state space in this case is effectively three-dimensional
and is illustrated in that reference. We note that the only
self-similar FRW models are the zero-curvature models which occur both
as an equilibrium point in the plane symmetric invariant set (with
$\bqz=\pm1=-2\bqp$), and as an orbit in the interior of the state
space.

\subsection{Monotone function}

The evolution equation for $\Xi\equiv\bqp+k\bu$ is given by
\begin{eqnarray}
  \Xi'&=&\bqz\Xi\left[-2(1-\bqp^2-\bu^2)+
  \frac{\gamma}{1+(\gamma-1)v^2}\Omega_n\right]
  \nonumber \\ &&
  -\frac{1}{2}\frac{(2-\gamma)+(3\gamma-2)v^2}{1+(\gamma-1)v^2}\Omega_n ,
\end{eqnarray}
which we note is of the same form as the evolution equation for $\bqp$
in the perfect fluid case. Noting the form of the constraint
(\ref{eq:sss-constr}), we consider the following function $\bar{M}$
(cf. \cite{GNU1}): 
\begin{eqnarray}
  \bar{M}&=&(1-\bqz^2)^{2-\gamma}\Xi^{3\gamma-4}\bc^{-\gamma}v^2
 (1-v^2)^{-(2-\gamma)} \label{eq:monosss}.
\end{eqnarray}
A direct calculation then yields:
\begin{eqnarray}
  \bar{M}'&=&-\left[\frac{(3\gamma-2)(2-\gamma)}{\gamma v}(1-v^2)\bc\right]
  \bar{M} ,
\end{eqnarray}
that is, $\bar{M}$ is monotonic in both the $v<0$ and the $v>0$ regions. 

When $v=0$ the constraint yields $\Xi\bc=0$. But since $\bc=0$, $v=0$
is an invariant (boundary) set, if $\bc=0$ then $\bc=0$, $v=0$
always. Hence on the surface $v=0$ and in the interior of the state
space $\bc>0$. Setting $v=0$ in the evolution equation for $v$ then
gives 
\begin{eqnarray}
  v'&=&-\frac{2-\gamma}{\gamma}\bc ,
\end{eqnarray}
which is strictly negative. That is, all orbits in the interior region
pass from $v>0$ to $v<0$ across the surface $v=0$ (i.e. the surface
$v=0$ acts as a membrane). Consequently there can be no closed or
recurrent orbits in the interior of the state space.

\section{Equilibrium points for the spatially self-similar
  case}\label{sec:ssseq} 

\begin{table}
  \caption{Summary of possible attractors for the spatially
    self-similar case.}\label{tab:sss}
  \begin{center}
    \begin{tabular}{|l|l|}\hline\hline
      K-rings $_\pm{\rm K}$ & 
      subset of $_+{\rm K}$ ($_-{\rm K}$) is always sources (sinks) \\
      Scalar-field dominated $_\pm\Phi$ & 
      $_+\Phi$ ($_-\Phi$) sink (source) when 
      $\gamma<\frac{4}{3}$, $k^2>1$ ($\kappa^2<2$) \\ \hline\hline
    \end{tabular}
  \end{center}
\end{table}

We shall display all of the equilibrium points below along with their 
eigenvalues. We will not present the corresponding eigenvectors explicitly.
In what follows, $\sqz=\pm1$ is the sign of $\bqz$, which indicates
whether the corresponding solution is expanding $(+)$ or contracting $(-)$.
The quantity $\bar{\omp}=\bu^2+\bw^2$ is the scalar-field contribution to 
the density parameter $\Omega_n$. The order of the dependent variables is 
$\left(\bqz,\bqp,\bc,v,\bu,\bw\right)$. The '$\pm$' suffices on the labels 
for equilibrium points correspond to the sign of $\bqz$ (i.e. the
value of $\sqz$). Equilibrium points that act as attractors are listed
in table \ref{tab:sss}.

\subsection{No scalar field ($\bu=0$, $\bw=0$)}

\subsubsection{K-points\\}

These are special points on the K-rings, defined in subsection
\ref{sec:Kasnerrings}. They all have $\bqz=\pm1$, $\bqp=\pm1$, and all
other variables equal to zero.

\subsubsection{Flat Friedmann\\}

$_\pm{\rm F}$: $\left(\sqz,-\frac{1}{2}\sqz,0,0,0,0\right)$.\\
$\Omega_n=\frac{3}{4}$, $\bar{\omp}=0$, 
$q_{\rm pf}=\frac{3\gamma-2}{2}$. \\
Eigenvalues ($\bc$ eliminated):
\begin{eqnarray}
  &&
  \frac{1}{2}(3\gamma-2)\sqz , \quad
  -\frac{3}{4}(2-\gamma)\sqz , \quad
  \frac{1}{4}(3\gamma-2)\sqz , \quad
  -\frac{3}{4}(2-\gamma)\sqz , \quad
  \frac{3\gamma}{4}\sqz .
  \nonumber
\end{eqnarray}
These points are saddles.

\subsubsection{Self-similar Kantowski-Sachs\\}

The state space contains the non-self-similar Kantowski-Sachs
solutions as a boun\-dary submanifold. In this submanifold, the
self-similar Kantowski-Sachs solution appears as an equilibrium
point.\\ 
$_\pm{\rm KS}$: $\left(\sqrt{-\frac{2-\gamma}{4(\gamma-1)}}\sqz,
-\sqrt{-\frac{\gamma-1}{2-\gamma}}\sqz,0,0,0,0\right)$, $\gamma<2/3$. \\
As this solution is physical only when $\gamma<2/3$, we will not
consider it further.

\subsection{Massless scalar field ($\bu\neq0$, $\bw=0$)}

\subsubsection{K-rings\\}\label{sec:Kasnerrings}

$_\pm{\rm K}$: 
$\left(\sqz,\pm\sqrt{1-\bu^2},0,0,\bu,0\right)$.\\
$\Omega_n=0$, $\bar{\omp}=\bu^2$, $q_{\rm pf}=2$. \\
Eigenvalues ($\bc$ eliminated):
\begin{eqnarray}
  &&
  2\sqz , \quad
  (2-\gamma)\left[\bqp+2\sqz\right] , \quad
  2(\gamma-1)\sqz+\gamma\bqp , \quad
  0 , \quad
  2\sqz+\bqp-\frac{\bu}{k} .
  \nonumber
\end{eqnarray}
Each K-ring corresponds to a one-parameter family of equilibrium points 
(and hence gives rise to a zero eigenvalue). They are analogues of the 
Kasner solutions in the case with no scalar field \cite{GNU1}.
For each K-ring, there is a subset of future or past attractors.\\
$_+{\rm K}$: sources and saddles.\\
$_-{\rm K}$: sinks and saddles.

\subsubsection{M-points\\}

$_\pm{\rm M}^{\tilde{v}}$: 
$\left(\sqz,-k^2f\sqz,f,\tilde{v}\sqz,kf\sqz,0\right)$, 
$f=1/(1+k^2)$, $\gamma>\frac{6}{5}$,\\
$\tilde{v}_{1,2}=\frac{-\gamma\left[(\gamma-1) - 
\frac{2-\gamma}{\gamma}k^2\right]\pm
  \sqrt{(\gamma-1)(2-\gamma)(3\gamma-2)+
    \gamma^2\left[(\gamma-1)-\frac{2-\gamma}{\gamma}k^2\right]^2}}
{(\gamma-1)(3\gamma-2)}$.\\
$\Omega_n=0$, $\bar{\omp}=\left(\frac{k}{1+k^2}\right)^2$.\\
Eigenvalues ($\bc$ eliminated):
\begin{eqnarray}
  &&
  -2f\sqz , \quad
  -2(1-k^2)f\sqz , \quad
  -(1-k^2)f\sqz , \quad
  F_1(\gamma,k)f\sqz ,
  \nonumber \\ &&
  -\frac{1}{\gamma v}
  \left[2-\gamma + (3\gamma-2)v^2+2\gamma v\right]f\sqz , \quad
  \nonumber
\end{eqnarray}
These equilibrium points are related to the Milne points $\tilde{\rm M}$
in \cite{GNU1}. They are only physical ($|\tilde{v}|<1$) for certain 
ranges of $\gamma$ and $k$. For instance, when $k<1$ it follows that 
$|\tilde{v}_2|>1$. When $k>1$, these points are saddles. Furthermore, 
examining the eigenvalues numerically for $0<k<1$, it turns out that 
the points always are saddles.

\subsubsection{Curvature-scaling solutions\\}

$_\pm{\rm X}^{\hat{v}}$: 
$\left(2kg\sqz,-kg\sqz,g,\hat{v}\sqz,g\sqz,0\right)$, 
$g=1/(\sqrt{2}\sqrt{1+k^2})$, $k<1$ ($\kappa^2>2$),\\
$\hat{v}_{1,2}=\frac{-(3\gamma-4)\gamma\frac{k}{2}\pm
  \sqrt{(\gamma-1)(2-\gamma)(3\gamma-2)+(3\gamma-4)^2\gamma^2\frac{k^2}{4}}}
{(\gamma-1)(3\gamma-2)}$.\\
$\Omega_n=0$, $\bar{\omp}=\frac{1}{2(1+k^2)}$.\\
Eigenvalues ($\bc$ eliminated):
\begin{eqnarray}
  &&
  -\frac{1-k^2}{k}g\sqz , \quad
  F_2(\gamma,k,\hat{v})\sqz, \quad
  F_3(\gamma,k,\hat{v})\sqz,
  \nonumber \\ &&
  F_4(\gamma,k,\hat{v})\sqz, \quad
  F_5(\gamma,k,\hat{v})\sqz. 
  \nonumber
\end{eqnarray}
For $k=1$ ($\kappa^2=2$) these points coincide with 
$_\pm{\rm M}^{\tilde{v}}$, and for $k>1$ ($\kappa^2<2$) they are unphysical. 
Note that $0<\hat{v}_1<1$ and $-1<\hat{v}_2<0$ 
only when $\gamma>4/3$, assuming $k<1$. For $\gamma<4/3$, it follows that 
$|\hat{v}|>1$, and the equilibrium points are unphysical. Numerical 
evaluation of the eigenvalues shows that these points are saddles.

\subsubsection{Equilibrium lines with arbitrary $v$\\}

$_\pm\Phi^v$: $\left(\sqz,-2\frac{\gamma-1}{\gamma}\sqz,0,v,
\pm\frac{1}{\gamma}\sqrt{(2-\gamma)(3\gamma-2)},0\right)$. \\
$\Omega_n=0$, $\bar{\omp}=(2-\gamma)(3\gamma-2)/\gamma^2$.\\
Eigenvalues ($\bqp$ eliminated):
\begin{eqnarray}
  &&
  0 , \quad
  0 , \quad
  2\sqz , \quad
  \frac{2}{\gamma}(2-\gamma)\sqz , \quad
  \frac{2}{\gamma}\sqz-\frac{\bu}{k}.
  \nonumber
\end{eqnarray}
There are two zero eigenvalues for these points. The first zero
eigenvalue corresponds to the fact that we have a line of
equilibrium points. The second zero eigenvalue indicates that the
equilibria are non-hyperbolic. For $v=0,\pm1$, these equilibrium lines
coincide with the various K-rings, see subsections \ref{sec:Kasnerrings}
and \ref{sec:Kasnerrings-tilt}, and these exceptional points mark where 
the K-rings change stability. The higher-order zero eigenvalue 
of $_\pm\Phi^v$ corresponds to the eigenvalue associated with the fact that 
$_\pm{\rm K}$ is a line of equilibrium points (and not to the eigenvalue that 
becomes zero due to the stability change of $_\pm{\rm K}$), and the 
corresponding eigenvector is 
$\vec{v}=\frac{\bu}{\bqp}\vec{e}_\bqp+\vec{e}_\bu$.
Perturbing the equilibrium lines $_\pm\Phi^v$ along this eigenvector, 
we find that
\begin{eqnarray}
  \bqp'&=&-2(1-\bqp^2-\bu^2)\sqz\bqp , \label{eq:lbqp} \\
  \bu' &=&-2(1-\bqp^2-\bu^2)\sqz\bu \label{eq:lbu} .
\end{eqnarray}
This is precisely the dynamical system restricted to the invariant set
$\bqz=\sqz$, $\bc=0$, $\bw=0$, $\Omega_n=0$. We can explicitly
integrate equations (\ref{eq:lbqp}) and (\ref{eq:lbu}). It follows that
$\bqp$ is proportional to $\bu$, and the orbits in the ($\bqp,\bu$)
plane consist of straight lines through the origin with additional
equilibrium points at $_\pm\Phi^v$ (where
$\bqp=-2\frac{\gamma-1}{\gamma}\bqz$), which are thus non-linear saddles.

\subsection{Scalar field with potential ($\bu\neq0$, $\bw\neq0$)}

There are a number of solutions with a non-zero potential
listed below. There are also equilibrium points $_\pm{\rm Z}^{v^*}$ 
with variable values\\
$\left(\sqz,0,\frac{1}{\sqrt{1-k^2}},v^*,0,
\sqrt{\frac{-2k^2}{1-k^2}}\right)$,
$v^*_{1,2}=\frac{-(\gamma-1)\gamma\sqrt{1-k^2}\sqz\pm
\sqrt{(\gamma-1)(2-\gamma)(3\gamma-2)+(\gamma-1)^2\gamma^2(1-k^2)}}
{(\gamma-1)(3\gamma-2)}$, \\
but these points are unphysical, since either $\bc$ or 
$\bw$ is imaginary.

\subsubsection{Scalar-field dominated solutions\\}

$_\pm\Phi$: 
$\left(\sqz,-\frac{1}{2}\sqz,0,0,
\frac{1}{2k}\sqz,\frac{1}{2k}\sqrt{3k^2-1}\right)$, 
$k^2>\frac{1}{3}$ ($\kappa^2<6$).\\
$\Omega_n=0$, $\bar{\omp}=\frac{3}{4}$, 
$q_{\rm pf}=\frac{1}{k^2}(1-k^2)=-\frac{1}{2}(2-\kappa^2)$. \\ 
Eigenvalues (constraint degenerate):
\begin{eqnarray}
  &&
  -\frac{k^2-1}{k^2}\sqz , \quad
  -\frac{3k^2-1}{2k^2}\sqz , \quad
  -\frac{3k^2-1}{2k^2}\sqz , \quad
  -\frac{k^2-1}{2k^2}\sqz ,
  \nonumber \\ &&
  \frac{3\gamma-4}{2}\sqz , \quad
  -\frac{3\gamma k^2-2}{2k^2}\sqz .
  \nonumber
\end{eqnarray}
When $\Omega_n=0$, the constraint defines two hypersurfaces
($\bc=0$ and $\bqp+k\bu=0$), and these hypersurfaces coincide at
$_\pm\Phi$. It turns out that the constraint is degenerate ($\nabla G=0$) at
these equilibrium points. Consequently, all eigenvector directions are
physical there, so we need to keep all six eigenvalues.
For $k^2=1/3$ ($\kappa^2=6$) these points coincide with points in the 
K-rings, and for $k^2<1/3$ ($\kappa^2>6$) they are unphysical.\\
$_+\Phi$: sink when $\gamma<\frac{4}{3}$, $k^2>1$ ($\kappa^2<2$); saddle 
otherwise.\\
$_-\Phi$: source when $\gamma<\frac{4}{3}$, $k^2>1$ ($\kappa^2<2$); 
saddle otherwise.\\
We note that $_+\Phi$ corresponds to the flat FRW power-law
inflationary solution \cite{Wetterich1,Wetterich2}.

\subsubsection{Curvature-scaling solutions\\}

$_\pm\Xi$: $\left(\frac{1}{2kg}\sqz,
-kg\sqz,0,0,g\sqz,\frac{1}{\sqrt{2}}\right)$,
$g=1/(\sqrt{2}\sqrt{1+k^2})$, 
$k>1$ ($\kappa^2<2$). \\
$\Omega_n=0$, $\bar{\omp}=\frac{2+k^2}{2(1+k^2)}$, 
$q_{\rm pf}=\frac{1-k^2}{2+k^2}=\frac{\kappa^2-2}{2(\kappa^2+1)}<0$. \\
Eigenvalues (constraint degenerate):
\begin{eqnarray}
  &&
  -\frac{1}{k}\left[2(\gamma-1)+\gamma k^2\right]\,g\sqz , \quad
  -\frac{1}{2k}\left[k^2+1\pm\sqrt{(k^2+1)(9k^2-7)}\right]\,g\sqz
  , \nonumber \\
  &&
  -\frac{1}{k}(k^2+1)\,g\sqz , \quad
  -\frac{1}{k}(k^2-1)\,g\sqz , \quad
  \frac{1}{k}\left[\gamma-(2-\gamma)(k^2+1)\right]\,g\sqz .
  \nonumber
\end{eqnarray}
For $k=1$ ($\kappa^2=2$) these points coincide with $_\pm\Phi$, and
for $k<1$ ($\kappa^2>2$) they are unphysical. When physical, these 
points are always saddles.

\subsubsection{Friedmann scaling solution\\}

$_\pm{\rm FS}$: $\left(\sqz,-\frac{1}{2}\sqz,0,0,
\frac{3k}{4}\gamma\sqz,\frac{3k}{4}\sqrt{\gamma(2-\gamma)}\right)$,
$k^2<\frac{2}{3\gamma}$ ($\kappa^2>3\gamma$).\\
$\Omega_n=\frac{3}{8}(2-3\gamma k^2)$, $\bar{\omp}=\frac{9}{8}\gamma k^2$,
$q_{\rm pf}=\frac{3\gamma-2}{2}$. \\
Eigenvalues ($\bc$ eliminated):
\begin{eqnarray}
  &&
  \frac{1}{2}(3\gamma-2)\sqz , \quad
  -\frac{3}{4}(2-\gamma)\sqz , \quad
  1 ,
  \nonumber \\ &&
  -\frac{3}{8}\left[(2-\gamma)
  \pm\sqrt{(2-\gamma)(12\gamma^2k^2-9\gamma+2)}\right]\sqz .
  \nonumber
\end{eqnarray}
For $k^2=\frac{2}{3\gamma}$ ($\kappa^2=3\gamma$) these points coincide with
$_\pm\Phi$, and for $k^2>\frac{2}{3\gamma}$ ($\kappa^2<3\gamma$) they are
unphysical. When physical, these points are always saddles.

\section{Timelike self-similar case}\label{sec:tss}

In the timelike self-similar case, the line element can be written
\cite{Bogoyavlensky} 
\begin{eqnarray}
  ds^2&=&e^{2t}d\tilde{s}^2 , \\
  d\tilde{s}^2&=&-D_1^2dt^2+dx^2+D_2^2d\Omega^2 , \\
  D_1&=&e^{\beta^0-2\beta^+} , \quad 
  D_2=e^{\beta^0+\beta^+} .
\end{eqnarray}
The kinematic quantities of the congruence normal to the symmetry surfaces
are related to ($\beta^0$, $\beta^+$) by:
\begin{equation}
  \theta=3\dot{\beta}^0 , \quad 
  \sigma_+=3\dot{\beta}^+ .
\end{equation}
Note that for the timelike self-similar case, the symmetry surfaces $x=$ 
constant are timelike, so the normal congruence is spacelike.
As for the spatial case, it is convenient to boost the kinematic quantities
in order to simplify the constraint \cite{GNU2}:
\begin{equation}
  \theta=\frac{1}{\sqrt{3}}\left(2\bth+\bsi\right) , \quad
  \sigma_+=\frac{1}{\sqrt{3}}\left(\bth+2\bsi\right) .
\end{equation}
The evolution equations for the metric functions $B_1\equiv D_1^{-1}$ and 
$B_2\equiv D_2^{-1}$ are
\begin{eqnarray}
  \dot{B_1}&=&\frac{1}{\sqrt{3}}\bsi B_1 , \\
  \dot{B_2}&=&-\frac{1}{\sqrt{3}}(\bth+\bsi)B_2 .
\end{eqnarray}
In terms of the coordinates we use, the physical quantities associated
with the perfect fluid are given by 
\begin{eqnarray}
  u_{\rm pf}^a&=&\frac{e^{-t}}{\sqrt{1-u^2}}
  \left(e^{2\beta^+-\beta^0},u,0,0\right) ,\\
  \mu&=&\frac{1-u^2}{1+(\gamma-1)u^2}e^{-2t}\mu_t ,
\end{eqnarray}
where $u$ is related to the tilt, and $\mu_t$ is the energy density with 
respect a congruence projected onto the surfaces of symmetry.
The conservation equations for the perfect fluid yield
\begin{eqnarray}
  \dot{u}&=&\frac{1-u^2}{\sqrt{3}\gamma\left[u^2-(\gamma-1)\right]}
  \left\{\gamma\left[2(\gamma-1)\bth+\gamma\bsi\right]u
  \right. \nonumber \\ && \left.
  +\sqrt{3}\left[(\gamma-1)(3\gamma-2)-(2-\gamma)u^2\right]B_1\right\} .
\end{eqnarray}
There is also an evolution equation for $\dot{\mu}_t$, but as it is
rather lengthy and not used elsewhere, we will not give it here. The
conservation equation for the scalar field is
\begin{eqnarray}
  \dot{X}&=&-\frac{1}{\sqrt{3}}(2\bth+\bsi)X
  +2kB_1^2-\frac{{\cal V}}{k} ,
\end{eqnarray}
and the Einstein field equations give:\\
{\em The Friedmann equation}:
\begin{eqnarray}
  \mu_t&=&\frac{1}{3}\frac{1+(\gamma-1)u^2}{u^2+(\gamma-1)}
  \left(\bth^2-\bsi^2-3(1+k^2)B_1^2-3B_2^2-3X^2+3{\cal V}\right) 
  \label{eq:FriedmannTSS}.
\end{eqnarray}
{\em Constraint equation}:
\begin{eqnarray}
  0&=&\gamma u\mu_t-\frac{2}{\sqrt{3}}
  \left[1+(\gamma-1)u^2\right](\bsi+\sqrt{3}kX)B_1 .
\end{eqnarray}
{\em Evolution equations for $\bth$ and $\bsi$}:
\begin{eqnarray}
  \dot{\bth}&=&-\frac{1}{\sqrt{3}}\left(\bth^2+\bsi^2+\bth\bsi-3(1+k^2)B_1^2
  +3X^2+3{\cal V}
  \right. \nonumber \\ && \left.
  -\frac{3(\gamma-1)(1-u^2)}{1+(\gamma-1)u^2}\mu_t\right) , \\
  \dot{\bsi}&=&-\frac{1}{\sqrt{3}}\left(\bsi^2+2\bth\bsi+6k^2B_1^2-3{\cal V}
  \right. \nonumber \\ && \left.
  +\frac{3}{2}\frac{(3\gamma-2)+(2-\gamma)u^2}{1+(\gamma-1)u^2}\mu_t \right) .
\end{eqnarray}

From (\ref{eq:FriedmannTSS}) and by demanding $\mu_t\geq0$ it follows that
\begin{eqnarray}
  \ty&=&\sqrt{\bth^2+3{\cal V}} 
\end{eqnarray}
is a dominant quantity. Consequently, we define bounded variables as
follows: 
\begin{eqnarray}
  &&\tqz=\frac{\bth}{\ty} , \quad
  \tqp=\frac{\bsi}{\ty} , \quad
  \tc=\frac{\sqrt{3}B_1}{\ty} , \quad
  \tct=\frac{\sqrt{3}B_2}{\ty} , \quad 
  \tu=\frac{\sqrt{3}X}{\ty} .
\end{eqnarray}
The Friedmann equation becomes an equation for the density parameter 
$\Omega_t$:
\begin{eqnarray}
  \Omega_t&=&\frac{3\mu_t}{\ty^2} \\
  &=&\frac{1+(\gamma-1)u^2}{u^2+(\gamma-1)}
  \left(1-\tqp^2-(1+k^2)\tc^2-\tu^2-\tct^2\right) ,
\end{eqnarray}
while the constraint becomes
\begin{eqnarray}
  0&=&\gamma u\Omega_t-2\left[1+(\gamma-1)u^2\right](\tqp+k\tu)\tc .
\end{eqnarray}
By defining a new independent variable
\begin{eqnarray}
  '&=&\frac{d}{d\xi}=\frac{\sqrt{3}}{\ty}\frac{d}{dx} ,
\end{eqnarray}
the evolution equation for $\ty$
\begin{eqnarray}
  \ty'&=&-\left\{\tqz\left[1+\tqp^2+\tqz\tqp-(1+k^2)\tc^2+\tu^2
  \right. \right. \nonumber \\ && \left. \left.
  -\frac{(\gamma-1)(1-u^2)}{1+(\gamma-1)u^2}\Omega_t\right]
  +\frac{1}{k}\tu(1-\tqz^2)\right\}\ty 
\end{eqnarray}
decouples, and we obtain the following reduced set of evolution equations:
\begin{eqnarray}
  \tqz'&=&-(1-\tqz^2)\left[1+\tqp^2+\tqz\tqp-(1+k^2)\tc^2+\tu^2-
  \frac{1}{k}\tqz\tu-\right.\nonumber \\ &&\left.
  \frac{(\gamma-1)(1-u^2)}{1+(\gamma-1)u^2}\Omega_t\right] , \\
  \tqp'&=& -\tqp\left\{(\tqz+\tqp)(1-\tqz\tqp)
  -\frac{1}{k}\tu(1-\tqz^2)
  \right. \nonumber \\ && \left. 
  +\tqz\left[(1+k^2)\tc^2-\tu^2+
  \frac{(\gamma-1)(1-u^2)}{1+(\gamma-1)u^2}\Omega_t\right]
  \right\}
  \nonumber \\ && 
  -\left[2k^2\tc^2-(1-\tqz^2)+\frac{1}{2}
  \frac{(3\gamma-2)+(2-\gamma)u^2}{1+(\gamma-1)u^2}\Omega_t\right] , \\
  \tc' &=&  \tc\left\{\tqz+\tqp+\tqz\left[\tqp^2+\tqz\tqp-(1+k^2)\tc^2+\tu^2-
  \right.\right.\nonumber \\ &&\left.\left.
  \frac{(\gamma-1)(1-u^2)}{1+(\gamma-1)u^2}\Omega_t\right]+
  \frac{1}{k}\tu(1-\tqz^2)\right\} , \\
  u'   &=& \frac{1-u^2}{\gamma\left[u^2-(\gamma-1)\right]}
  \left\{\gamma\left[2(\gamma-1)\tqz+\gamma\tqp\right]u
  \right. \nonumber \\ && \left.
  +\left[(\gamma-1)(3\gamma-2)-(2-\gamma)u^2\right]\tc\right\} ,
  \label{eq:u} \\
  \tu' &=& \tu\left\{-(\tqz+\tqp)+\tqz\left[\tqp^2+\tqz\tqp-(1+k^2)\tc^2+
  \tu^2-  \right.\right.\nonumber \\ &&\left.\left.
  \frac{(\gamma-1)(1-u^2)}{1+(\gamma-1)u^2}\Omega_t\right]\right\}+
  2k\tc^2-\frac{1}{k}(1-\tqz^2)(1-\tu^2) , \\
  \tct'&=&  \tct\left\{-\tqp+\tqz\left[\tqp^2+\tqz\tqp-(1+k^2)\tc^2+\tu^2-
  \right.\right.\nonumber \\ &&\left.\left.
  \frac{(\gamma-1)(1-u^2)}{1+(\gamma-1)u^2}\Omega_t\right]+
  \frac{1}{k}\tu(1-\tqz^2)\right\} .
\end{eqnarray}
This system is invariant under the transformation
\begin{equation}
  (\xi,\tqz,\tqp,\tc,u,\tu,\tct)\rightarrow
  (-\xi,-\tqz,-\tqp,\tc,-u,-\tu,\tct) .
\end{equation}
Furthermore, noting the invariance under $(\tc,u)\rightarrow(-\tc,-u)$ 
and under $\tct\rightarrow-\tct$, we can without loss of 
generality restrict the analysis to $\tc\geq0, \tct\geq0$.

Note that the denominator of the evolution equation for $u$
(\ref{eq:u}) is zero when $u=\pm\sqrt{\gamma-1}$. The only way to pass
this sonic hypersurface without introducing a shock wave is when the
numerator also is zero. This defines a submanifold of the sonic
hypersurface, and this submanifold is only physical for the
$u=-\sqrt{\gamma-1}$ case. This severely restricts the global dynamics
\cite{GNU2}.

\subsection{Invariant submanifolds}

A number of invariant submanifolds can be identified:
\begin{itemize}
  \item Massless scalar field: $V=0$, which implies $\tqz=\pm1$.
  \item Static: $\tc=0,u=0$ (see \cite{GNU2}).
  \item Plane-symmetric: $\tct=0$ (see \cite{GNU2}).
  \item No perfect fluid: $\Omega_t=0$ (and $u$ decouples).
  \item No scalar field: $\tu=0$, $\tqz=\pm1$ and $k=0$.
\end{itemize}

The global dynamics of the submanifold with no scalar field has been
studied previously in \cite{GNU2}, where state-space diagrams can be
found.

\subsection{Monotone function}

As for the SSS case, it is possible to generalise the monotone
function for the perfect fluid TSS case \cite{GNU2} to the case of a
perfect fluid with a scalar field. This is done by replacing
$\bar{\Sigma}_+$ with $\tqp+k\tu$ in equation (32) of \cite{GNU2}. Thus,
\begin{eqnarray}
  \tilde{M}&=&(\tqp+k\tu)^{3\gamma-4} \tc^{-\gamma} u^{-2(\gamma-1)}
  (1-u^2)^{-(2-\gamma)} \tct^{2(2-\gamma)}, \label{eq:monotss}
\end{eqnarray}
with
\begin{equation}
  \tilde{M}'=\left[\frac{(3\gamma-2)(2-\gamma)}{\gamma u}(1-u^2)\tc\right]
  \tilde{M},
\end{equation}
is monotonic in the regions $u>0$ and $u<0$. Furthermore,
\begin{equation}
  u'|_{u=0}=-\frac{3\gamma-2}{\gamma}\tc ,
\end{equation}
which is strictly negative. Consequently, as in the SSS case, there
can be no closed or recurrent orbits in the interior of the state
space.

\section{Equilibrium points for the timelike self-similar
  case}\label{sec:tsseq} 

\begin{table}
    \caption{Summary of possible attractors for the timelike
      self-similar case.}\label{tab:tss}
  \begin{center}
    \begin{tabular}{|l|l|}\hline\hline
      K-rings $_\pm{\rm K}$ & 
      subset of $_+{\rm K}$ ($_-{\rm K}$) is always sources (sinks) \\
      Scalar-field dominated $_\pm\Phi$ & 
      $_+\Phi$ ($_-\Phi$) source (sink) when \\
      & $\gamma<\frac{4}{3}$, $k^2<2\frac{\gamma-1}{\gamma}$
      ($\kappa^2>\frac{\gamma}{\gamma-1}$) \\ \hline\hline
    \end{tabular}
  \end{center}
\end{table}

We shall display all of the equilibrium points below along with their
eigenvalues. We will not present the corresponding eigenvectors
explicitly. In what follows, $\sqz=\pm1$ is the sign of $\tqz$, which
indicates whether the corresponding solution is expanding $(+)$ or
contracting $(-)$. The order of the dependent variables is
$\left(\tqz,\tqp,\tc,u,\tu,\tct\right)$. The '$\pm$' suffices on the
labels for equilibrium points correspond to the sign of $\tqz$
(i.e. the value of $\sqz$). The quantity $\tilde{\omp}=\tu^2+\tqz^2-1$
indicates the presence of a non-zero scalar field. Equilibrium points
that act as attractors are listed in table \ref{tab:tss}.

\subsection{No scalar field ($\tu=0$, $V=0$)}

The state space contains a number of solutions with no scalar field, as 
presented below. There is also a solution with variable values
$\left(\sqz,-\frac{3\gamma-2}{4(\gamma-1)}\sqz,
0,0,0,0\right)$, $\Omega_t=-\frac{(2-\gamma)(7\gamma-6)}{16(\gamma-1)^3}$.
As this solution is physical only when $6/7<\gamma<1$, we will not
consider it further.

\subsubsection{K-points\\}

These are special points on the K-rings, defined in section
\ref{sec:KasnerringsTSS}. They all have $\tqz=\pm1$, $\tqp=\pm1$, and
all other variables equal to zero.

\subsubsection{Static solution\\}

$_\pm{\rm T}$: $\left(\sqz,-2\frac{\gamma-1}{3\gamma-2}\sqz,0,0,0,
\frac{\sqrt{\gamma^2+4(\gamma-1)}}{3\gamma-2}\right)$.\\
$\Omega_t=\frac{4(\gamma-1)}{(3\gamma-2)^2}$, $\tilde{\omp}=0$. \\
Eigenvalues ($\tc$ eliminated):
\begin{eqnarray}
  &&
  -\sqz, \quad
  \frac{2-\gamma}{3\gamma-2}\sqz, \quad
  \frac{2\gamma}{3\gamma-2}\sqz, \quad
  -\frac{1}{2}\sqz\pm\frac{\sqrt{\gamma^2-44\gamma+36}}{2(3\gamma-2)}\sqz.
  \nonumber
\end{eqnarray}
These points are always saddles.

\subsubsection{Regular centre\\}

$_\pm{\rm C}^0$: $\left(\sqz,0,0,0,0,1\right)$.\\
$\Omega_t=0$, $\tilde{\omp}=0$.\\
Eigenvalues (constraint degenerate):
\begin{eqnarray}
  &&
  -2\sqz, \quad
  -\sqz, \quad
  -\sqz, \quad
  \sqz, \quad
  2\sqz, \quad
  2\sqz.
  \nonumber
\end{eqnarray}
These points are always saddles.

\subsection{Massless scalar field ($\tu\neq0$, $V=0$)}

\subsubsection{K-rings\\}\label{sec:KasnerringsTSS}

$_\pm{\rm K}$: $\left(\sqz,\pm\sqrt{1-\tu^2},0,0,\tu,0\right)$.\\
$\Omega_t=0$, $\tilde{\omp}=\tu^2$.\\
Eigenvalues ($\tc$ eliminated):
\begin{eqnarray}
  &&
  \sqz, \quad
  -2\sqz-\frac{\gamma}{\gamma-1}\tqp, \quad
  4\sqz+2\tqp-\frac{2}{k}\tu, \quad
  0, \quad
  4\sqz+\frac{3\gamma-2}{\gamma-1}\tqp.
  \nonumber
\end{eqnarray}
Each K-ring corresponds to a one-parameter family of equilibrium points 
(and hence gives rise to a zero eigenvalue). They are analogues of the 
Kasner solutions in the case with no scalar field \cite{GNU2}. 
For each K-ring, there is a subset of future or past attractors.\\
$_+{\rm K}$: sources and saddles.\\
$_-{\rm K}$: sinks and saddles.

\subsubsection{M-points\\}

$_\pm{\rm M}^{\tilde{u}}$: 
$\left(\sqz,-k^2f\sqz,f,\tilde{u}\sqz,kf\sqz,0\right)$, 
$f=1/(1+k^2)$, $\gamma>\frac{6}{5}$, \\
$\tilde{u}_{1,2}=\frac{\gamma(\gamma-1) - \frac{2-\gamma}{2}\gamma k^2 \pm
  \sqrt{(\gamma-1)(2-\gamma)(3\gamma-2)+
    \gamma^2\left[(\gamma-1)-\frac{2-\gamma}{2} \gamma k^2\right]^2}}
{2-\gamma}$.\\
$\Omega_t=0$, $\tilde{\omp}=f^2k^2$.\\
Eigenvalues ($\tc$ eliminated):
\begin{eqnarray}
  &&
  -2f\sqz, \quad
  -(1-k^2)f\sqz, \quad
  -2(1-k^2)f\sqz, \quad
  F_6(\gamma,k,u)\sqz ,
  \nonumber \\ &&
  -\frac{1}{\gamma u}
  \left[3\gamma-2 + (2-\gamma)u^2 +2\gamma u\right]f\sqz.
  \nonumber
\end{eqnarray}
These equilibrium points are related to the Milne points $\tilde{\rm M}$
in \cite{GNU2}. They are only physical ($|\tilde{u}|<1$) for certain 
ranges of $\gamma$ and $k$. For instance, when $k>1$ it follows that 
$|\tilde{u}_2|>1$. When $k>1$, these points are saddles. Furthermore, 
for $u=\tilde{u}_2$ they are saddles even when $k<1$. Examining the 
eigenvalues numerically for $0<k<1$, there are values of $\gamma$ and $k$ 
for which the $u=\tilde{u}_1$ points act as attractors. However, this only 
occurs when $\tilde{u}_1>\sqrt{\gamma-1}$, i.e., when the corresponding 
equilibrium point is beyond the sonic hypersurface located at 
$u=\sqrt{\gamma-1}$. As it is impossible to cross this sonic hypersurface
in a regular way, the $_+\tilde{M}$ points will not affect the dynamics of 
the models we are interested in, even though the points are attractors for 
some values of $\gamma$ and $k$. This situation is also present in the
case without a scalar field \cite{GNU2}.

\subsubsection{Curvature-scaling solutions\\}

$_\pm{\rm X}^{\hat{u}}$: $\left(\sqz,-\frac{1}{2}\sqz,\frac{1}{2k},
\hat{u}\sqz,\frac{1}{2k}\sqz,\frac{1}{\sqrt{2}\,k}\sqrt{k^2-1}\right)$, 
$k>1$ ($\kappa^2<2$),\\
$\hat{u}_{1,2}=\frac{\frac{3\gamma-4}{2}\gamma k \pm
  \sqrt{(\gamma-1)(2-\gamma)(3\gamma-2)+
  \left(\frac{3\gamma-4}{2}\right)^2\gamma^2k^2}}
{2-\gamma}$.\\
$\Omega_t=0$, $\tilde{\omp}=\frac{1}{4k^2}$.\\
Eigenvalues ($\tc$ eliminated):
\begin{eqnarray}
  &&
  \frac{k^2-1}{k^2}\sqz, \quad
  -\frac{1}{2k}\left[k\pm\sqrt{4-3k^2}\right]\sqz, \quad
  F_7(\gamma,k,u) , 
  \nonumber \\ &&
  -\left[1+\frac{3\gamma-2+(2-\gamma)u^2}{2\gamma k u}\right]\sqz.
  \nonumber
\end{eqnarray}
For $k=1$ ($\kappa^2=2$) these points coincide with 
$_\pm{\rm M}^{\tilde{u}}$, and for $k<1$ ($\kappa^2>2$) they are unphysical. 
Noting that $k>1$, these points are always saddles.

\subsubsection{Equilibrium lines with arbitrary $u$\\}

$_\pm\Phi^u$: $\left(\sqz,-2\frac{\gamma-1}{\gamma}\sqz,0,u,
\pm\frac{1}{\gamma}\sqrt{(2-\gamma)(3\gamma-2)},0\right)$. \\
$\Omega_t=0$, $\tilde{\omp}=(2-\gamma)(3\gamma-2)/\gamma^2$.\\
Eigenvalues ($\tqp$ eliminated):
\begin{eqnarray}
  &&
  0, \quad
  0, \quad
  \sqz, \quad
  \frac{2}{\gamma}(2-\gamma)\sqz, \quad
  \frac{4}{\gamma}\sqz-\frac{2\tu}{k}.
  \nonumber
\end{eqnarray}
There are two zero eigenvalues for these points. The first zero
eigenvalue corresponds to the fact that we have a line of equilibrium
points. The second zero eigenvalue indicates that the equilibria are
non-hyperbolic. For $u=0,\pm1$, these equilibrium lines coincide with
the various K-rings (see subsections \ref{sec:KasnerringsTSS} and
\ref{sec:Kasnerrings-tilt}) and these exceptional points mark where
the K-rings change stability. The higher-order zero eigenvalue of
$_\pm\Phi^u$ corresponds to the one that indicates that $_\pm{\rm K}$
is a line of equilibrium points (and not to the one that becomes zero
due to the stability change of $_\pm{\rm K}$). The corresponding
eigenvector is $\vec{v}=\frac{\tu}{\tqp}\vec{e}_\tqp+\vec{e}_\tu$.
Perturbing the equilibrium lines $_\pm\Phi^u$ along this eigenvector,
we find that 
\begin{eqnarray}
  \tqp'&=&-2(1-\tqp^2-\tu^2)\sqz\tqp , \label{eq:ltqp} \\
  \tu' &=&-2(1-\tqp^2-\tu^2)\sqz\tu \label{eq:ltu} .
\end{eqnarray} 
This is precisely the dynamical system restricted to the invariant set 
$\tqz=\sqz$, $\tc=0$, $\tct=0$, $\Omega_t=0$. We can explicitly
integrate equations (\ref{eq:ltqp}) and (\ref{eq:ltu}). It follows that
$\tqp$ is proportional to $\tu$, and the orbits in the ($\tqp,\tu$)
plane consist of straight lines through the origin with additional
equilibrium points at $_\pm\Phi^u$ (where
$\tqp=-2\frac{\gamma-1}{\gamma}\tqz$), which are thus non-linear saddles.

\subsection{Scalar field with potential ($\tu\neq0$, $V\neq0$)}

There is a number of solutions with non-zero potential
listed below. There are also equilibrium points $_\pm{\rm \Xi}$ 
with variable values 
$\left(\frac{1}{h}\sqz,-k^2h\sqz,0,0,kh\sqz,\sqrt{1-k^2}\right)$,
$h=1/\sqrt{1+k^2}$, $k^2<1$ ($\kappa^2>2$), but these points are 
unphysical since $|\tqz|>1$ when $k>0$.

\subsubsection{Scalar-field dominated solutions\\}

$_\pm\Phi$: 
$\left(2kh\sqz,-kh\sqz,0,0,h\sqz,0\right)$, 
$k^2<\frac{1}{3}$ ($\kappa^2>6$), $h=1/\sqrt{1+k^2}$.\\
$\Omega_t=0$, $\tilde{\omp}=(2-3k^2)h^2$.\\
Eigenvalues (constraint degenerate):
\begin{eqnarray}
  &&
  -\frac{3\gamma-4}{\gamma-1}kh\sqz, \quad
  \frac{1-3k^2}{k}h\sqz, \quad
  \frac{1-3k^2}{k}h\sqz, \quad
  \frac{1-k^2}{k}h\sqz, \quad
  \frac{1-k^2}{k}h\sqz,
  \nonumber \\ &&
  \frac{2(\gamma-1)-\gamma k^2}{(\gamma-1)k}h\sqz.
  \nonumber
\end{eqnarray}
As in the timelike case, since the constraint is degenerate we must retain
all six eigenvalues.
For $k=1/3$ ($\kappa^2=6$), these points coincide with points in the
K-rings, and for $k^2>\frac{1}{3}$ ($\kappa^2<6$) they are unphysical.\\
$_+\Phi$: source when $\gamma<4/3$ and
$k^2<2\frac{\gamma-1}{\gamma}$ ($\kappa^2>\frac{\gamma}{\gamma-1}$);
saddle otherwise.\\
$_-\Phi$: sink when $\gamma<4/3$ and
$k^2<2\frac{\gamma-1}{\gamma}$ ($\kappa^2>\frac{\gamma}{\gamma-1}$);
saddle otherwise.

\subsubsection{Potential-dominated solutions\\}

$_\pm{\rm Z}^{u^*}$:
$\left(\sqrt{\frac{1-k^2}{1+k^2}}\sqz,0,\frac{1}{\sqrt{1+k^2}},u^*,0,0\right)$,
$k<1$ $(\kappa^2>2)$,\\
$u^*_{1,2}=\frac{(\gamma-1)\gamma\sqrt{1-k^2}\sqz\pm
\sqrt{(\gamma-1)(2-\gamma)(3\gamma-2)+(\gamma-1)^2\gamma^2(1-k^2)}}
{2-\gamma}$. \\
$\Omega_t=0$, $\tilde{\omp}=\frac{2k^2}{1+k^2}$.\\
Eigenvalues ($\tc$ eliminated):
\begin{eqnarray}
  &&
  -\tqz, \quad
  -(1\pm\sqrt{5})\tqz, \quad
  F_8(\gamma,k),
  \nonumber \\ &&
  \frac{2(1-u^2)+\gamma(u^2-2\sqrt{1+k^2}\tqz u-3)}
  {\gamma u \sqrt{1+k^2}}.
  \nonumber
\end{eqnarray}
For these solutions, the potential is non-zero, since $|\tqz|<1$. In the 
context of this paper, these solutions may be unphysical. They are on the 
boundary and correspond to non-self-similar solutions with a cosmological 
constant. They are always saddles.

\section{Equilibrium points at extreme tilt}\label{sec:extreme}

In addition to the above equilibrium points, there is a number of
equilibrium points for which the tilt is extreme, i.e., $v=\pm1$ or
$u=\pm1$. These are artifacts of the particular approach that we have
adopted, and signify that the coordinates break down. These points are
still important since orbits that are asymptotic to them may pass
between the spacelike and the timelike self-similar regions (at least
at non-vacuum equilibrium points). Furthermore, for submanifolds where
the tilt variable is not specified (e.g. the fluid vacuum
submanifold), some of these solutions are indistinguishable from
similar ones with non-extreme tilt. Equilibrium points in this class
that act as attractors are listed in table \ref{tab:tilt}. In what
follows, the equilibrium points will be given both in the SSS
variables $(\bqz,\bqp,\bc,v,\bu,\bw)$ and in the TSS variables
$(\tqz,\tqp,\tc,u,\tu,\tct)$. 

\begin{table}
    \caption{Summary of possible attractors with extreme
      tilt.}\label{tab:tilt} 
  \begin{center}
    \begin{tabular}{|l|l|}\hline\hline
      K-rings $_\pm{\rm K}^\pm$ & 
      subset of $_+{\rm K}$ ($_-{\rm K}$ ) is always sources (sinks) \\
      M-points $_\pm{\rm M}^\pm$ & $_+{\rm M}^\pm$ ($_-{\rm M}^\pm$) sinks
      (sources) when \\
      & $\gamma>\frac{2(3+k^2)}{5+k^2}$, $k<1$ ($\kappa^2>2$) \\ 
      H-lines $_\pm{\rm H}^\mp$ & 
      subset of $_+{\rm H}^-$ ($_-{\rm H}^+$) is always sinks (sources) \\
      Scalar-field dominated $_\pm\Phi^\pm$ & 
      SSS: $_+\Phi^\pm$ ($_-\Phi^\pm$) sinks (sources) when \\
      & $\gamma>\frac{4}{3}$, $k>1$ ($\kappa^2<2$) \\  
      & TSS: $_+\Phi^\pm$ ($_-\Phi^\pm$) sources (sinks) when \\
      & $\gamma<\frac{4}{3}$, $k^2<\frac{1}{3}$ ($\kappa^2>6$) \\ \hline\hline
    \end{tabular}
  \end{center}
\end{table}

\subsection{No scalar field}

\subsubsection{K-points\\}

These are special points on the K-rings, defined in 
section \ref{sec:Kasnerrings-tilt}.
They all have $\bqz=\tqz=\pm1$, $\bqp=\tqp=\pm1$, $v=u=\pm1$, and all
other variables equal to zero.

\subsubsection{C points\\}

In the TSS case there are equilibrium points that resemble the regular
centre $_\pm{\rm C}^0$, but have $u=\pm1$.\\ 
$_\pm{\rm C}^\pm$:
$\left(\sqz,0,0,\pm1,0,1\right)$ (TSS). \\
$\Omega_t=0$, $\tilde{\omp}=0$.  \\
Eigenvalues ($\tct$ eliminated):
\begin{eqnarray}
  &&
  \sqz, \quad
  -\sqz, \quad
  -4\frac{\gamma-1}{2-\gamma}\sqz, \quad
  \frac{\sqz\pm3}{2}.
  \nonumber
\end{eqnarray}
These points are always saddles.

\subsection{Massless scalar field}

\subsubsection{K-rings\\}\label{sec:Kasnerrings-tilt}

$_+{\rm K}^\pm$, $_-{\rm K}^\pm$: 
$\left(\sqz,\pm\sqrt{1-U^2},0,\pm1,U,0\right)$ (SSS and TSS).\\
$\Omega=0$, $\omp=U^2$. \\
Eigenvalues (SSS: $\bc$ eliminated, TSS: $\tc$ eliminated):
\begin{eqnarray}
  {\rm SSS} &:&2\sqz , \quad
  2(\sqz+\bqp) , \quad
  -2\frac{2(\gamma-1)\sqz+\gamma \bqp}{2-\gamma} ,
  \nonumber \\ &&
  0 , \quad
  2\sqz+\bqp-\frac{\bu}{k} .
  \nonumber \\
  {\rm TSS} &:&\sqz , \quad
  2(\sqz+\tqp) , \quad
  -2\frac{2(\gamma-1)\sqz+\gamma \tqp}{2-\gamma} , \quad
  \nonumber \\ &&
  0 , \quad
  2\left(2\sqz+\tqp-\frac{\tu}{k}\right) .
  \nonumber
\end{eqnarray}
For each K-ring, there is a subset of future or past attractors.\\
$_+{\rm K}^\pm$: sources and saddles.\\
$_-{\rm K}^\pm$: sinks and saddles.

\subsubsection{M-points\\}

a) $_\pm{\rm M}^\pm$:
$\left(\sqz,-k^2f\sqz,f,\sqz,kf\sqz,0\right)$,
$f=1/(1+k^2)$ (SSS and TSS). \\
$\Omega=0$, $\omp=\left(\frac{k}{1+k^2}\right)^2$. \\
Eigenvalues (SSS: $\bc$ eliminated, TSS: $\tc$ eliminated):
\begin{eqnarray}
  &&
  -2f\sqz , \quad
  -4f\sqz , \quad
  -2\frac{(5+k^2)\gamma-2(3+k^2)}{2-\gamma}f\sqz ,
  \nonumber \\ &&
  -2(1-k^2)f\sqz , \quad
  -(1-k^2)f\sqz .
  \nonumber 
\end{eqnarray}
$_+{\rm M}^+$: sink when $\gamma>\frac{2(3+k^2)}{5+k^2}$, $k<1$
($\kappa^2>2$); saddle otherwise.\\
$_-{\rm M}^-$: source when $\gamma>\frac{2(3+k^2)}{5+k^2}$, $k<1$
($\kappa^2>2$); saddle otherwise.

b) $_\pm{\rm M}^\mp$:
$\left(\sqz,-k^2f\sqz,f,-\sqz,kf\sqz,0\right)$,
$f=1/(1+k^2)$ (SSS and TSS). \\
$\Omega=0$, $\omp=\left(\frac{k}{1+k^2}\right)^2$. \\
Eigenvalues (SSS: $\bc$ eliminated, TSS: $\tc$ eliminated):
\begin{eqnarray}
  &&
  -2f\sqz , \quad
  -(1-k^2)f\sqz , \quad
  -2(1-k^2)f\sqz , \quad
  -2(1-k^2)f\sqz , \quad
  0 .
  \nonumber
\end{eqnarray}
The zero eigenvalue is due to the fact that these points are the end
points of the equilibrium lines $_\pm{\rm H}$ (see section \ref{sec:H}).\\
$_+{\rm M}^-$:  sink when $k<1$ ($\kappa^2>2$); saddle otherwise.\\
$_-{\rm M}^+$:  source when $k<1$ ($\kappa^2>2$); saddle otherwise.

\subsubsection{H-lines\\}\label{sec:H}

$_\pm{\rm H}^\mp$:
$\left(\sqz,Q_+,1+\sqz Q_+,-\sqz,k(1+\sqz Q_+)\sqz,0\right)$,
$-1<\sqz Q_+<-\frac{k^2}{1+k^2}$ (SSS and TSS).\\
$\Omega=-2\sqz(1+\sqz Q_+)\left[(1+k^2)Q_+ +\sqz k^2\right]$,
$\Omega_\phi=k^2(1+\sqz Q_+)^2$.\\
Eigenvalues (SSS: $\bc$ eliminated, TSS: $\tc$ eliminated):
\begin{eqnarray}
  &&
  -2(1+2\sqz Q_+)\sqz, \quad
  -2(1+2\sqz Q_+)\sqz, \quad
  -(1+2\sqz Q_+)\sqz,
  \nonumber \\ &&
  -2(1+\sqz Q_+)\sqz, \quad
  0 .
  \nonumber
\end{eqnarray}
These equilibria consist of lines of equilibrium points. The
eigenvector direction along the lines is
$\vec{v}=\vec{e}_{Q_+}+k\vec{e}_U$. The end points of the lines are
the M-points $_\pm{\rm M}^\mp$ at one end and points of stability
change on the K-rings $_\pm{\rm K}^\mp$ at the other end.\\
$_+{\rm H}^-$:  always contains at least a subset of sinks. Solely
sinks when $k>1$ ($\kappa^2<2$); sources and saddles otherwise.\\
$_-{\rm H}^+$:  always contains at least a subset of sources. Solely
sources when $k>1$ ($\kappa^2<2$); sources and saddles otherwise.

\subsubsection{Curvature-scaling solutions\\}

$_+{\rm X}^{\pm}$, $_-{\rm X}^{\pm}$:
$\left(2kg\sqz,-kg\sqz,g,\pm1,g\sqz,0\right)$, \\
$g=1/(\sqrt{2}\sqrt{1+k^2})$, $k<1$ ($\kappa^2>2$) (SSS), \\
$\left(\sqz,-\frac{1}{2}\sqz,\frac{1}{2k},\pm1,
\frac{1}{2k}\sqz,\frac{1}{\sqrt{2}\,k}\sqrt{k^2-1}\right)$,
$k>1$ ($\kappa^2<2$) (TSS). \\
$\Omega=0$, $\omp=\frac{1}{2(1+k^2)}$. \\
Eigenvalues (SSS: $\bc$ eliminated, TSS: $\tc$ eliminated):
\begin{eqnarray}
  {\rm SSS} &:&-2\frac{3\gamma-4}{2-\gamma}\left[\sgn{v}+k\right]g\sqz , \quad
  -\frac{1-k^2}{k}g\sqz , \quad
  F_9(k)\sqz,
  \nonumber \\ &&
  F_{10}(k)\sqz, \quad
  -F_{11}(k,v) ,
  {\rm where \ } F_10(k)>0, {\rm \ and} \quad F_{11}(k)>0 .
  \nonumber \\
  {\rm TSS} &:&
  \frac{1}{k^2}(k^2-1)\sqz , \quad
  -\frac{1}{k}\left[k\sqz+\sgn{v}\right] , \quad
  \nonumber \\ &&
  -\frac{1}{k}\left[k\sqz+\sgn{v}\right]\frac{(3\gamma-4)}{2-\gamma}, \quad
  -\frac{1}{2}\sqz\pm\frac{1}{2k}\sqrt{4-3k^2}) .
  \nonumber
\end{eqnarray}
For $k=1$ ($\kappa^2=2$) these points coincide with $_\pm{\rm M}^\pm$.
They are always saddles.

\subsection{Scalar field with potential ($U\neq0$, $W\neq0$)}

\subsubsection{Scalar-field dominated solutions\\}

$_+\Phi^\pm$, $_-\Phi^\pm$: $\left(\sqz,-\frac{1}{2}\sqz,0,\pm1,
\frac{1}{2k}\sqz,\frac{1}{2k}\sqrt{3k^2-1}\right)$,
$k^2>\frac{1}{3}$ ($\kappa^2<6$) (SSS), \\
$\left(2kh\sqz,-kh\sqz,0,\pm1,h\sqz,0\right)$,
$h=1/\sqrt{1+k^2}$, $k^2<\frac{1}{3}$ ($\kappa^2>6$) (TSS). \\
$\Omega=0$, $\omp=3/4$.  \\
Eigenvalues (SSS: $\bqp$ eliminated, TSS: $\tqp$ eliminated):
\begin{eqnarray}
  {\rm SSS} &:&
  -\frac{k^2-1}{k^2}\sqz , \quad
  -\frac{k^2-1}{2k^2}\sqz , \quad
  -\frac{3\gamma-4}{2-\gamma}\sqz , \quad
  -2\bw^2\sqz, \quad
  -2\bw^2\sqz .
  \nonumber \\
  {\rm TSS} &:&
  \frac{h}{k}(1-3k^2) , \quad
  \frac{h}{k}(1-3k^2) , \quad
  \frac{h}{k}(1-k^2) , \quad
  \frac{h}{k}(1-k^2) ,
  \nonumber \\ &&
  -2\frac{3\gamma-4}{2-\gamma}hk .
\nonumber 
\end{eqnarray}
For $k^2=1/3$ ($\kappa^2=6$) these points coincide with points on the
K-rings $_\pm{\rm K}^\pm$.\\
SSS: \\
$_+\Phi^\pm$: sinks when $\gamma>\frac{4}{3}$, $k>1$ ($\kappa^2<2$); 
saddles otherwise.\\
$_-\Phi^\pm$: sources when $\gamma>\frac{4}{3}$, $k>1$ ($\kappa^2<2$); 
saddles otherwise.\\
TSS: \\
$_+\Phi^\pm$: sources when $\gamma<\frac{4}{3}$, $k^2<1/3$
($\kappa^2>6$); saddles otherwise. \\ 
$_-\Phi^\pm$: sinks when $\gamma<\frac{4}{3}$, $k^2<1/3$
($\kappa^2>6$); saddles otherwise.

\subsubsection{Curvature-scaling solutions\\}

$_+{\rm \Xi}^\pm$, $_-{\rm \Xi}^\pm$: 
$\left(\frac{1}{2kg}\sqz,-kg\sqz,0,\pm1,g\sqz,\frac{1}{\sqrt{2}}\right)$,
$g=1/(\sqrt{2}\sqrt{1+k^2})$, 
$k>1$ ($\kappa^2<2$) (SSS), \\
$\left(\frac{1}{h}\sqz,-k^2h\sqz,0,\pm1,kh\sqz,\sqrt{1-k^2}\right)$,
$h=1/\sqrt{1+k^2}$,
$k<1$ ($\kappa^2>2$) (TSS). \\ 
$\Omega=0$, $\omp=\frac{2+k^2}{2(1+k^2)}$. \\
Eigenvalues (SSS: $\bqp$ eliminated, TSS: $\tqp$ eliminated):
\begin{eqnarray}
  {\rm SSS} &:&
  -\frac{\sqrt{k^2+1}\pm\sqrt{9k^2-7}}{2\sqrt{2}k}\sqz , \quad
  -\frac{1}{k}(k^2-1)g\sqz , \quad
  -\frac{1}{2kg}\sqz ,
  \nonumber \\ & & 
  -\frac{2}{(2-\gamma)k}\left[2(\gamma-1)+(3\gamma-2)k^2\right]g\sqz .
  \nonumber \\
  {\rm TSS} &:&
  -\frac{1}{h}\sqz , \quad
  (1-k^2)h\sqz , \quad
  -\frac{1}{2h}\sqz\pm\frac{1}{2}\sqrt{7-9k^2} ,
  \nonumber \\ & & 
  -2\frac{2(\gamma-1)-(2-\gamma)k^2}{2-\gamma}h\sqz .
  \nonumber
\end{eqnarray}
For $k=1$ ($\kappa^2=2$) these points coincide with $_\pm\Phi^\pm$. 
They are always saddles.

\subsubsection{Potential-dominated solutions\\}

$_\pm{\rm Z}^\pm$:
$\left(\sqrt{\frac{1-k^2}{1+k^2}}\sqz,0,\frac{1}{\sqrt{1+k^2}},\pm1,
0,0\right)$, $k<1$ $(\kappa^2>2)$.\\
$\Omega_t=0$, $\omp=\frac{2k^2}{1+k^2}$.\\
Eigenvalues ($\tc$ eliminated):
\begin{eqnarray}
  &&
  -\tqz, \quad
  -(1\pm\sqrt{5})\tqz, \quad
  -2\frac{\sqrt{1+k^2}\tqz +\sgn{u}}
  {\sqrt{1+k^2}}, 
  \nonumber \\ & & 
  -2\frac{(3\gamma-4)\sgn{u}+2(\gamma-1)\sqrt{1+k^2}\tqz}
  {(2-\gamma)\sqrt{1+k^2}} .
  \nonumber
\end{eqnarray}
These solutions are only physical in the TSS region. 
They are always saddles.

\section{Global results and discussion}\label{sec:global}

Due to the existence of monotone functions and the fact that there are
consequently no closed or periodic orbits in the physical state spaces
we can obtain global results for the dynamics by studying the local
stability of the equilibria. 

Indeed, from the monotone functions obtained in the spatially
self-similar (SSS) case (\ref{eq:monosss}) and the timelike
self-similar (TSS) case (\ref{eq:monotss}) we can immediately deduce
from the {\em monotonicity principle} \cite{WE} that all orbits have
$\bqz^2\rightarrow1$, $\bqp+k\bu\rightarrow0$ or $\bc\rightarrow0$
(or an extreme value for $v$) asymptotically in the SSS case (and
similarly in the TSS case). Moreover, we can also see immediately that
by setting the right-hand-sides of equations (\ref{eq:bc}) and
(\ref{eq:bw}) to zero that either $\bc=0$ or $\bw=0$, or if both are
non-zero then $k\bqp+\bu=0$; this latter case yields very severe
constraints on any possible equilibrium points. In fact, from the
local analysis of the equilibria we can determine all of the sinks and
sources. In both the SSS and TSS cases a set of massless scalar field
models lying on the $_+{\rm K}$-ring act as sources (i.e., early-time
attractors) and a set of massless scalar field models lying on the
$_-{\rm }$K-ring act as sinks (i.e., late-time attractors), and for
certain ranges of the parameters (e.g., $\kappa^2<2$)  the equilibrium
point $_+\Phi$ with $U\ne0$ and $V\ne0$, corresponding to the
ever-expanding inflationary flat FRW power-law solution
\cite{Wetterich1,Wetterich2,ColeyIbanezVanDenHoogen,bchio}, are sinks
(i.e., late-time attractors). [We note that the equilibrium point
$_-\Phi$ corresponding to $\sqz=-1$, which acts as a source,
represents an ever-contracting solution and therefore is of less
physical importance, although it does serve to classify all of the
possible orbits in the state space.] Hence we have the global results
that models that are initially expanding always expand from an initial
singularity and always recollapse to a second singularity
(when $\kappa^2>2$) or either recollapse or expand forever towards a
flat FRW power-law solution (for $\kappa^2<2$). This global behaviour
is the same as that for positive curvature FRW models and
Kantowski-Sachs models \cite{CG-KS} and for Bianchi type IX models
\cite{R9,R92}. Models that expand from an initial singularity and
recollapse to a second singularity are said to satisfy the
positive-curvature recollapse property \cite{LW1,LW2}. Models that expand
towards the flat FRW power-law solution isotropise and inflate to the
future and are said to satisfy the cosmic no-hair theorem
\cite{R9,R92}. The time-reverse of the above dynamics is also possible
(essentially $\sqz\rightarrow-\sqz$; although we have included this in
the analysis these models are of less interest physically).
Solutions in which the shear and the kinetic energy of the scalar field
dominate are analogues of the Kasner and Jacobs solutions \cite{WE},
and a rigorous study of the structure of the singularity, which is
non-oscillatory, for a general class of analytic solutions of the
Einstein field equations coupled to a massless scalar field has
recently been presented \cite{AR}.

However, there are some aspects of the global dynamics of the
self-similar, spherically symmetric models that are different.
The complete set of attractors for different values of the parameters
$\gamma$ and $\kappa$ are summarised in table \ref{tab:total}. Some of
these differences are quite subtle. First, we note that the flat FRW
power-law inflationary solution corresponds both to a set of
non-tilted equilibrium points $_\pm\Phi$ and to points at extreme tilt
$_\pm\Phi^\pm$. There is a bifurcation of the equation-of-state
parameter at $\gamma=4/3$ in that the power-law inflationary solution
$\Phi$ is a non-tilted attractor $_\pm\Phi$ for $\gamma<4/3$ and an
extreme-tilt attractor $_\pm\Phi^\pm$ when $\gamma>4/3$. For
$\gamma=4/3$ there exist lines of equilibrium points with arbitrary
tilt $v$ (or $u$ in the TSS case). This type of $\gamma$-dependent
behaviour has also been found in Bianchi type V two-fluid models
\cite{G,GN}. Second, there exist additional  M-point attractors
(for $\gamma>\frac{2(3\kappa^2+2)}{5\kappa^2+2}$, $\kappa^2>2$)
at extreme tilt. The significance of these is less clear, although
they are important for the matching of orbits and they are related to
critical phenomena \cite{ccgnu-letter}. We shall discuss this further
in sections \ref{sec:fluidvac} and \ref{sec:massless}. We recall that
solutions in the SSS region and the TSS region can be matched across
the surface of extreme tilt via the equilibrium points with extreme
tilt (see earlier and \cite{GNU1,GNU2}). In addition, a comprehensive
analysis of the matching of solutions would be necessary in order to
obtain a complete knowledge of the intermediate dynamics of the
models. Clearly the intermediate behaviour of the models under
investigation will be quite different to that of the models previously
studied. 

We note that all of the equilibrium points with non-negligible matter,
namely the non-vacuum Flat Friedmann $_\pm{\rm F}$ and the Friedmann
Scaling $_\pm{\rm FS}$ equilibrium points, are {\em saddles}. This means 
that the perfect fluid is not dynamically important asymptotically. In order 
to understand the asymptotic behaviour of the models we consequently
need only study the vacuum models (i.e., the invariant boundary with
$\Omega_n=0$ in the SSS case and the invariant boundary with
$\Omega_t=0$ in the TSS case). We shall discuss the fluid vacuum
models further in section \ref{sec:fluidvac}. The matter will play an
important role in describing the dynamics of the models at
intermediate times, and hence the physical properties of the
models. We shall illustrate some of the intermediate dynamics in the
next two sections. However, in the SSS case we know from the behaviour
of the monotone function that the subcase $\bc=0$ is important
asymptotically (in the TSS case the analogous case is $\tc=0$, leading
to the static models \cite{GNU2}). Moreover, when $\Omega_n=0$, the
constraint (\ref{eq:sss-constr}) leads to $(\bqp+k\bu)\bc=0$. Clearly
the invariant set $\bc=0$ (and $v=0$), corresponding to the
(non-self-similar) Kantowski-Sachs models, is of vital importance, and
knowledge of the dynamics of the Kantowski-Sachs models is crucial for
a complete understanding of the dynamics of the models under
consideration here. In addition, all of the interesting transient
dynamics with non-negligible matter (e.g., the non-vacuum Flat
Friedmann and the Friedmann Scaling saddle equilibrium points,  as
well as the power-law attractors) occurs in the Kantowski-Sachs
invariant submanifold. Consequently, we shall study the
Kantowski-Sachs models in more detail elsewhere \cite{CG-KS}. 

We note that a set of massless scalar field models lying on the K-rings 
act as sources and sinks (i.e., early- and late-time attractors). It is 
therefore also of interest to study the self-similar, spherically
symmetric massless scalar field models more fully. Indeed, such a
study will also be of relevance in the study of critical phenomena
(see section \ref{sec:massless}). We shall return to this in future
work. 

\begin{table}
    \caption{Summary of possible future attractors. Past attractors
      are obtained by changing the sign of $Q_0$, which is indicated
      by the lower-left index. This is also the sign of the
      expansion.
      Comments:
      (1) The attractors $_+{\rm H}^-$ have extreme tilt; orbits may
      pass between the SSS and TSS regions there. 
      (2) For $\gamma=4/3$ the SSS power-law solution switches from
      non-tilted $_+\Phi$ to $_+\Phi^\pm$ at extreme tilt; precisely at the
      bifurcation value there appear lines of equilibrium points between
      these two.
      (3) In the TSS case, the non-tilted $_-\Phi$ and the tilted
      $_-\Phi^\pm$ coexist for certain parameter values. However, they
      are always separated by the sonic hypersurface. 
      (4) In some regions further conditions must be fulfilled for some
      attractors:
      *) The attractor $_+{\rm M}^-$ must satisfy
      $\gamma>\frac{2(3\kappa^2+2)}{5\kappa^2+2}$, where (as $\kappa^2>2$)
      $\frac{6}{5}<\frac{2(3\kappa^2+2)}{5\kappa^2+2}<\frac{4}{3}$; 
      **) The non-tilted TSS attractor $_-\Phi$ must satisfy
      $\kappa^2>\frac{\gamma}{\gamma-1}$.}\label{tab:total}
  \begin{center}
    \begin{tabular}{|l|l|l|l|}\hline\hline
      & $1<\gamma<6/5$ & $6/5<\gamma<4/3$ & $4/3<\gamma<2$ \\ \hline

      & $_-$K-rings & $_-$K-rings & $_-$K-rings \\
      $\kappa^2<2$
      & $_+{\rm H}^-$-line & $_+{\rm H}^-$-line & $_+{\rm H}^-$-line \\ 
      & SSS power-law $_+\Phi$ & SSS power-law $_+\Phi$ &
      SSS power-law $_+\Phi^\pm$ \\ \hline

      & $_-$K-rings & $_-$K-rings & $_-$K-rings \\
      $2<\kappa^2<4$
      & $_+{\rm H}^-$-line & $_+{\rm H}^-$-line & $_+{\rm H}^-$-line \\ 
      & & SSS M-point $_+{\rm M}^-$ * & SSS M-point $_+{\rm M}^-$ \\ \hline

      & $_-$K-rings & $_-$K-rings & $_-$K-rings \\
      $4<\kappa^2<6$
      & $_+{\rm H}^-$-line & $_+{\rm H}^-$-line & $_+{\rm H}^-$-line \\ 
      & & SSS M-point $_+{\rm M}^-$ * & SSS M-point $_+{\rm M}^-$ \\
      & & TSS $_-\Phi$ ** & \\ \hline

      & $_-$K-rings & $_-$K-rings & $_-$K-rings \\
      & $_+{\rm H}^-$-line & $_+{\rm H}^-$-line & $_+{\rm H}^-$-line \\ 
      $\kappa^2>6$ & TSS $_-\Phi$ **
      & SSS M-point $_+{\rm M}^-$ * & SSS M-point $_+{\rm M}^-$ \\
      & TSS $_-\Phi^\pm$ & TSS $_-\Phi$ & \\
      & & TSS $_-\Phi^\pm$ & \\ \hline\hline
    \end{tabular}
  \end{center}
\end{table}

\section{Fluid vacuum}\label{sec:fluidvac}

When there is no barotropic fluid present ($\Omega_n=0$ and
$\Omega_t=0$, respectively), the constraint gives rise to two separate
invariant submanifolds: either $\bqp=-k\bu$ ($\tqp=-k\tu$), or else
$\bc=0$ ($\tc=0$). The $\bqp=-k\bu$ ($\tqp=-k\tu$) submanifold is
particularly interesting, as it contains all the sinks and sources of
the more general models under consideration. Furthermore, this
submanifold is three-dimensional, and hence lends itself to visual
presentation. In addition, the submanifold $\bc=0$ will be studied in
detail in \cite{CG-KS}.

\subsection{Spatially self-similar case}

In the SSS case, the reduced dynamical system becomes (eliminating the
variable $\bc$): 
\begin{eqnarray}
  \bqz'&=&(1-\bqz^2)\left[1-2(1+k^2)\bu^2\right] , \\
  \bu'&=&-2\left(\bqz\bu-\frac{k}{1+k^2}\right)\left[1-(1+k^2)\bu^2\right] 
  \nonumber \\ &&
  +\frac{1}{k}\left(1-\frac{2k^2}{1+k^2}\right)\bw^2 , \\
  \bw'&=&(1+k^2)\left(2\bqz\bu-\frac{1}{k}\right)\bu\bw .
\end{eqnarray}

\begin{figure}
  \centerline{\hbox{\epsfig{figure=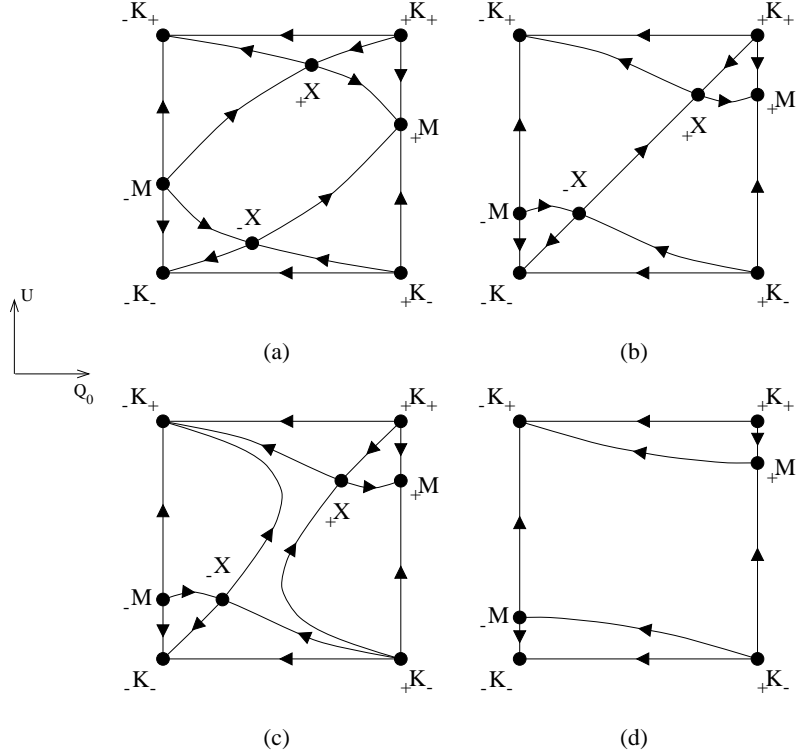,width=0.8\textwidth}}}
  \caption{The state space for the $\Omega_n=0$, $\bqp=-k\bu$, $\bw=0$ 
  submanifold for various values of $k$: (a) $k^2<1/3$ ($\kappa^2>6$); 
  (b) $k^2=1/3$ ($\kappa^2=6$); (c) $1/3<k^2<1$ ($2<\kappa^2<6$) and
  (d) $k>1$ ($\kappa^2<2$). The subscripts on the K-points refer to
  the sign of $\bqz$ and $\bu$, respectively. Note the separatrix cross-over 
  occurring for $k^2=1/3$ ($\kappa^2=6$).}\label{fig:sssvac-w0} 
\end{figure}

\begin{figure}
  \centerline{\hbox{\epsfig{figure=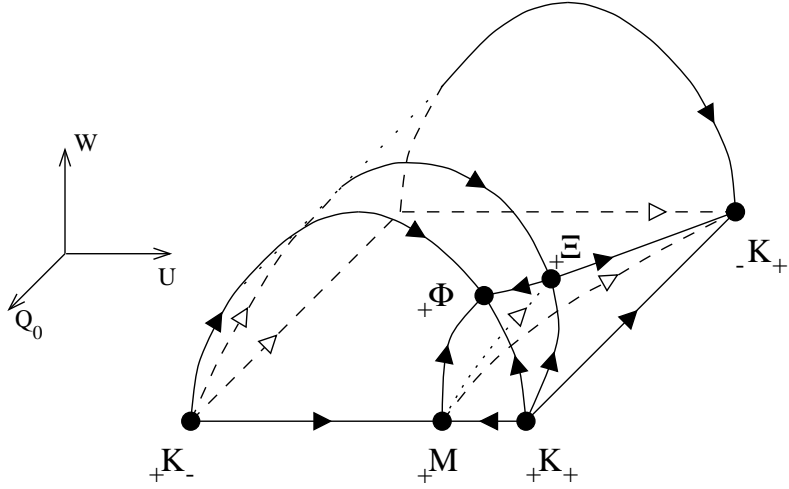,width=0.8\textwidth}}}
  \caption{The state space for the $\Omega_n=0$, $\bqp=-k\bu$ submanifold 
  for $k>1$ ($\kappa^2<2$). Dashed curves and white arrows
  are screened. The dotted curve between $_+{\rm M}$ and $_+\Xi$ is in the
  interior of the state space. The subscripts on the K-points refer to
  the sign of $\bqz$ and $\bu$, respectively. The state space is
  bounded by the 
  following invariant submanifolds: the bottom is the massless scalar field
  submanifold (note that only part of the bottom is depicted -- for a
  complete picture see Fig. \ref{fig:sssvac-w0}d), the half-disks are
  plane-symmetric submanifolds, and the half-cylinder is the
  Kantowski-Sachs submanifold ($\bc=0$). Note that orbits and
  equilibrium points in the rear half have been suppressed for
  clarity.}\label{fig:sssvac}   
\end{figure}

The equilibrium points of this system are listed in table
\ref{tab:sssvac}. They constitute a subset of the points 
listed in sections \ref{sec:ssseq} and \ref{sec:extreme}.
In figures \ref{fig:sssvac-w0} and \ref{fig:sssvac}, some examples of 
state-space diagrams for this model are displayed.

\begin{table}
    \caption{Equilibrium points in the $\Omega_n=0$, $\bqp=-k\bu$
      submanifold. Note that subscripts on labels have been
      suppressed.}\label{tab:sssvac}
  \begin{center}
    \begin{tabular}{|l|ccc|c|} \hline\hline
      & $\bqz$ & $\bu$ & $\bw$ &\\ \hline
      K & $\sqz$ & $\pm\frac{1}{\sqrt{1+k^2}}$ & 0 & \\
      M & $\sqz$ & $\frac{k}{1+k^2}\sqz$ & 0 & \\
      X & $\frac{2k}{\sqrt{2}\sqrt{1+k^2}}\sqz$ & \
      $\frac{1}{\sqrt{2}\sqrt{1+k^2}}\sqz$ & 0 & $k<1$ ($\kappa^2>2$)\\
      $\Phi$ & $\sqz$ & $\frac{1}{2k}\sqz$ & $\frac{1}{2k}\sqrt{3k^2-1}$ &
      $k^2>1/3$ ($\kappa^2<6$) \\
      $\Xi$ & $\frac{\sqrt{2}\sqrt{1+k^2}}{2k}\sqz$ &
      $\frac{1}{\sqrt{2}\sqrt{1+k^2}}\sqz$ & $\frac{1}{\sqrt{2}}$ &
      $k>1$ ($\kappa^2<2$) \\ \hline\hline
    \end{tabular}
  \end{center}
\end{table}

\subsection{Timelike self-similar case}

In the TSS case, the reduced dynamical system becomes (eliminating the
variable $\tc$): 
\begin{eqnarray}
  \tqz'&=& -(1-\tqz^2)\left[\tct^2-\frac{1+k^2}{k}(\tqz-2k\tu)\tu\right] ,\\
  \tu'&=&-\frac{1}{k}\left(1-\tqz^2+2k\tqz\tu-\frac{2k^2}{1+k^2}\right)
  \left[1-(1+k^2)\tu^2\right]
  \nonumber \\ &&
  +\left(\tqz\tu-\frac{2k}{1+k^2}\right)\tct^2 , \\
  \tct'&=&  \tct\left[(1-\tqz^2+2k\tqz\tu)\frac{1+k^2}{k}-
  \tqz(1-\tct^2)\right] .
\end{eqnarray}

\begin{figure}
  \centerline{\hbox{\epsfig{figure=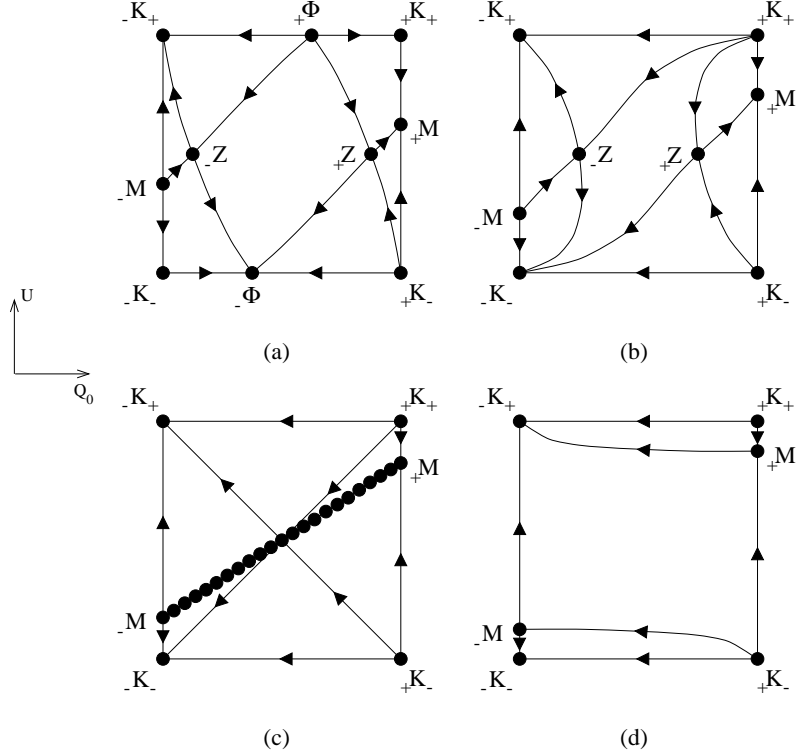,width=0.8\textwidth}}}
  \caption{The state space for the $\Omega_t=0$, $\tqp=-k\tu$, $\tct=0$ 
  submanifold for various values of $k$: (a) $k^2<1/3$ ($\kappa^2>6$); 
  (b) $1/3<k^2<1$ ($2<\kappa^2<6$); (c) $k=1$ ($\kappa^2=2$) and
  (d) $k>1$ ($\kappa^2<2$). The subscripts on the K-points refer to
  the sign of $\tqz$ and $\tu$, respectively. Note the line
  bifurcation occurring for $k=1$
  ($\kappa^2=2$).}\label{fig:tssvac-c20}  
\end{figure}

\begin{figure}
  \centerline{\hbox{\epsfig{figure=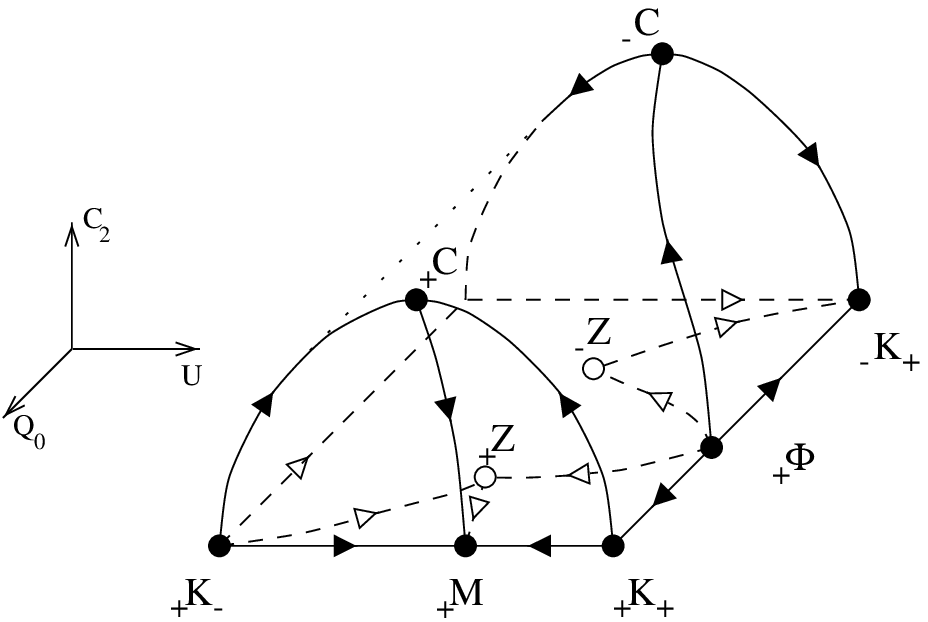,width=0.8\textwidth}}}
  \caption{The state space for the $\Omega_t=0$, $\tqp=-k\tu$ submanifold 
  for $k^2<1/3$ ($\kappa^2>6$). Dashed curves and white arrows and
  circles are screened. The subscripts on the K-points refer to
  the sign of $\tqz$ and $\tu$, respectively. The state space is
  bounded by the following invariant submanifolds: the bottom is the
  massless scalar field submanifold (note that only part of the bottom
  is depicted -- for a complete picture see
  Fig. \ref{fig:tssvac-c20}a), the half-disks are plane-symmetric
  submanifolds, and the half-cylinder is the static submanifold
  ($\tc=0$). Note that orbits and equilibrium points in the rear half
  have been suppressed for clarity.}\label{fig:tssvac} 
\end{figure}

The equilibrium points of this system are listed in table
\ref{tab:tssvac}. They constitute a subset of the points 
listed in sections \ref{sec:tsseq} and \ref{sec:extreme}.
In figures \ref{fig:tssvac-c20} and \ref{fig:tssvac}, some examples of 
state-space diagrams for this model are displayed.

\begin{table}
    \caption{Equilibrium points in the $\Omega_t=0$, $\tqp=-k\tu$
      submanifold. Note that subscripts on labels have been
      suppressed.}\label{tab:tssvac}
  \begin{center}
    \begin{tabular}{|l|ccc|c|} \hline\hline
      & $\tqz$ & $\tu$ & $\tct$ &\\ \hline
      C & $\sqz$ & 0 & 1 & \\
      K & $\sqz$ & $\pm\frac{1}{\sqrt{1+k^2}}$ & 0 & \\
      M & $\sqz$ & $\frac{k}{1+k^2}\sqz$ & 0 & \\
      X & $\sqz$ & $\frac{1}{2k}\sqz$ & $\frac{1}{\sqrt{2}\,k}\sqrt{k^2-1}$ & 
      $k>1$ ($\kappa^2<2$) \\
      $\Phi$ & $\frac{2k}{\sqrt{1+k^2}}\sqz$ &
      $\frac{1}{\sqrt{1+k^2}}\sqz$ & 0 & 
      $k^2<1/3$ ($\kappa^2>6$) \\
      $Z$ & $\sqrt{\frac{1-k^2}{1+k^2}}\sqz$ & 0 & 0 &
      $k<1$ ($\kappa^2>2$) \\ \hline\hline
    \end{tabular}
  \end{center}
\end{table}

\subsection{Discussion}

In both the SSS and the TSS cases there are always attractors
dominated by the kinetic part of the scalar field, corresponding to the 
K-equilibrium points. Consequently, there will always be solutions that expand
from a K-singularity and recollapse to a K-singularity. When
$\kappa^2>2$, the M-equilibrium points also are attractors. These
correspond to dispersing solutions. Finally, there are $\Phi$-equilibrium 
points, which correspond to power-law inflationary
solutions when $\kappa^2<2$. In the SSS case they act as attractors
when $\kappa^2<2$, while in the TSS case they are attractors when
$\kappa^2>6$. 

To summarise, when $\kappa^2<2$ the SSS case contains recollapsing K
$\rightarrow$ K solutions and power-law inflationary solutions K
$\rightarrow$ $\Phi$. This is the same situation as for the
Kantowski-Sachs models and closed Friedmann models examined in
\cite{CG-KS}. In contrast, when $\kappa^2<2$ the TSS case contains only
recollapsing solutions. When $\kappa^2>2$, both the SSS and the TSS
cases contain recollapsing solutions and also dispersing solutions K
$\rightarrow$ M. The asymptotics is thus similar to the fluid-only
case \cite{ccgnu}, where the generic regular solutions either are
recollapsing K $\rightarrow$ K solutions or ever-expanding solutions K
$\rightarrow$ M. Additionally, when $\kappa^2>6$ the TSS case also
contains dispersing K $\rightarrow$ $\Phi$ solutions that are
non-inflationary.

\section{Massless scalar field}\label{sec:massless}

Critical phenomena in gravitational collapse were first found by
Choptuik \cite{Choptuik} in the study of a massless scalar field, and
remain an active field of research (see, e.g., \cite{art:Gundlach1999}
and references therein). The solution at the threshold of black-hole
formation in spherically symmetric radiation fluid collapse,
corresponding to $\alpha=1/3$, was studied by Evans \& Coleman
\cite{art:EvansColeman1994}. In \cite{ccgnu-letter} a new class of
`asymptotically Minkowski' self-similar spacetimes were presented,
which were shown to be intimately related to the so-called critical
phenomena which arise in spherically symmetric gravitational collapse
calculations \cite{art:EvansColeman1994}. 

Here, we present the governing equations for a self-similar massless
scalar field in spherical symmetry. In this case, $\phi$ is still of
the form 
\begin{equation}
  \phi=\Phi(\xi)+\sqrt{2}\,k\eta,
\end{equation} 
and the corresponding equations are formally obtained by setting $V=0$.
In the presence of a barotropic fluid we then essentially have a
non-interacting {\it two-fluid} model \cite{CarrColey1999classi,CW},
in which the massless scalar field can be identified with a stiff
perfect fluid and the two fluids are separately conserved. We can then
deduce that the models evolve from the massless scalar field model to
the single-perfect fluid model \cite{CarrColey1999classi,CW}. 
Hereafter, we shall assume that there is no barotropic fluid present,
and that we are investigating a special case of the fluid vacuum model
studied above. The massless scalar field equations without perfect
fluid are then obtained by subsequently setting $\Omega_n=0$ and
$\Omega_t=0$, respectively. This leads to the decoupling of $v$ and
$\bc$ in the SSS case, and $u$ and $\tc$ in the TSS case,
respectively. Furthermore, in the SSS case $\bw=0$, while in the TSS
case, it follows that $\tqz=\sqz$ $(=\pm1)$. The remaining dynamical
systems are thus three-dimensional systems in $(\bqz,\bqp,\bu)$ and
$(\tqp,\tu,\tct)$, respectively. The constraint leads to two separate
regions: either $C_1=0$ or else $Q_+=-kU$. Note that the latter is a
special case of the models treated in section \ref{sec:fluidvac}.
Here, we briefly summarise the governing equations for these models
and list all of the equilibrium points.

\subsection{Spatially self-similar case}

In the SSS case, the Friedmann equation becomes
\begin{equation}
  0=1-\bqp^2-(1+k^2)\bc^2-\bu^2,
\end{equation}
which implies that
\begin{equation}    
  \bar{Z}\equiv1-\bqp^2-\bu^2 = (1+k^2)\bc^2 .
\end{equation}
The constraint becomes
\begin{equation}
  0=(\bqp+k\bu)\bar{Z} ,
\end{equation}
and the reduced dynamical system is
\begin{eqnarray}
  \bqz'&=&-(1-\bqz^2)(1-2\bar{Z}) , \\
  \bqp'&=&-2\bar{Z}\left(\bqz\bqp+\frac{k^2}{1+k^2}\right) ,\\
  \bu' &=&-2\bar{Z}\left(\bqz \bu-\frac{k  }{1+k^2}\right) .
\end{eqnarray}

The constraint implies that either $\bar{Z}$ or $\bqp+k\bu$ must be
zero. When $\bar{Z}=0$, both $\bqp$ and $\bu$ are constants (subject
to $1-\bqp^2-\bu^2=0$), whereby the dynamical system reduces to a
single evolution equation for $\bqz$, and consequently $\bqz$ is
monotonically decreasing. If the constant values of $\bqp$ and $\bu$
do not satisfy $\bqp+k\bu=0$, then the dynamics in the invariant set
$\bar{Z}=0$ does not intersect with the dynamics in the invariant set
$\bqp+k\bu=0$. The latter is contained within the fluid vacuum case
studied in the previous section (see figure \ref{fig:sssvac-w0}). The
equilibrium points of the system are listed in table
\ref{tab:sss-massless}. 

\begin{table}
    \caption{Equilibrium points in the $\Omega_n=0$, $\bw=0$
      submanifold. Note that subscripts on labels have been
      suppressed.}\label{tab:sss-massless}
  \begin{center}
    \begin{tabular}{|l|llll|}\hline\hline
      & $\bqz$ & $\bqp$ & $\bu$ & \\ \hline
      K-rings & $\sqz$, & \multicolumn{2}{c}{$\bqp^2+\bu^2=1$} & \\
      M & $\sqz$, & $-k^2f\sqz$ & $kf\sqz$ & $f=1/(1+k^2)$ \\
      X & $2kg\sqz$ & $-kg\sqz$ & $g\sqz$ &
      $g=1/(\sqrt{2}\sqrt{1+k^2})$ \\ \hline\hline
    \end{tabular}
  \end{center}
\end{table}

\subsection{Timelike self-similar case}

In the TSS case, the Friedmann equation becomes
\begin{equation}
  0=1-\tqp^2-(1+k^2)\tc^2-\tu^2-\tct^2,
\end{equation}
which implies that
\begin{equation}  
  \tilde{Z}\equiv 1-\tqp^2-\tu^2 = (1+k^2)\tc^2+\tct^2 .
\end{equation}
The constraint becomes
\begin{equation}
  0=(\tqp+k\tu)(\tct^2-\tilde{Z}) ,
\end{equation}
and the reduced dynamical system is
\begin{eqnarray}
  \tqp'&=&-\sqz\tqp(2\tilde{Z}-\tct^2)
  +\frac{2k^2}{1+k^2}(\tct^2-\tilde{Z}) , \\
  \tu' &=&-\sqz\tu(2\tilde{Z}-\tct^2)
  -\frac{2k}{1+k^2}(\tct^2-\tilde{Z}) , \\
  \tct'&=&\sqz\tct(1+\tct^2-2\tilde{Z}) .
\end{eqnarray}

The constraint implies that either $\tilde{Z}=\tct^2$ or
$\tqp=-k\tu$. When $\tilde{Z}=\tct^2$, the dynamical system becomes 
two-dimensional. The evolution equation for $\tct$ decouples and
$\tct$ and $\tqp$ (or $\tu$) are monotonic. If the values of $\tqp$
and $\tu$ do not satisfy $\tqp+k\tu=0$, then the dynamics in the
invariant set $\tilde{Z}=\tct^2$ does not intersect with the dynamics
in the invariant set $\tqp+k\tu=0$. The latter is contained within the
fluid vacuum case studied in the previous section, and corresponds to
the semi-disks at $\tqz=\pm1$ in figure \ref{fig:tssvac}. The
equilibrium points of the system are listed in table
\ref{tab:tss-massless}. 

\begin{table}
    \caption{Equilibrium points in the $\Omega_t=0$, $\tqz=\sqz$
      submanifold. Note that subscripts on labels have been
      suppressed.}\label{tab:tss-massless}
  \begin{center}
    \begin{tabular}{|l|llll|} \hline\hline
      & $\tqp$ & $\tu$ & $\tct$ & \\ \hline
      C & 0 & 0 & 1 & \\
      K-ring & \multicolumn{2}{c}{$\tqp^2+\tu^2=1$} & 0 & \\
      M & $-k^2f\sqz$ & $kf\sqz$ & 0 & $f=1/(1+k^2)$ \\ 
      X & $-\frac{1}{2}\sqz$ & $\frac{1}{2k}\sqz$ &
      $\pm\frac{1}{\sqrt{2}\,k}\sqrt{k^2-1}$ & \\ \hline\hline
    \end{tabular}
  \end{center}
\end{table}

\subsection{Discussion}

The dynamics is different in the various invariant submanifolds. The
$\bar{Z}=0$ submanifold of the SSS case only contains recollapsing
K $\rightarrow$ K solutions. The dynamics in the $\bqp=-k\bu$
submanifold of the SSS case is more complicated (see figure
\ref{fig:sssvac-w0}); when $\kappa^2<2$ there are only
recollapsing K $\rightarrow$ K solutions, when $2<\kappa^2<6$ there
are recollapsing solutions and dispersing K $\rightarrow$ M solutions,
and when $\kappa^2>6$ there are also singularity-free bouncing M
$\rightarrow$ M solutions, analogous to the bouncing Friedmann --
Lema\^{\i}tre solutions. The TSS case only contains dispersing
solutions. In the $\tilde{Z}=\tct^2$ submanifold there are K
$\rightarrow$ C solutions, while in the $\tqp=-\tu$ submanifold there
are K $\rightarrow$ M solutions.

\appendix

\section{Fluid quantities}\label{app:q}

For the non-tilted ($v=0$) equilibrium points of the SSS case, we have 
given the deceleration parameter $q_{\rm pf}$ with respect to the fluid 
congruence. The necessary expressions are summarised below.

The expansion of the fluid congruence is given by
\begin{eqnarray}
  \theta_{\rm pf}&=&\nabla_a u_{\rm pf}^a \nonumber \\
  &=&\frac{\by e^{-x}}{\sqrt{3}\sqrt{1-v^2}}
  \left(\frac{vv'}{1-v^2}+2\bqz+\bqp+3v\bc\right) ,
\end{eqnarray}
and the deceleration parameter, defined by
\begin{eqnarray}
  q_{\rm pf}&=&-\left(1+
  3\frac{u_{\rm pf}^a \nabla_a \theta_{\rm pf}}
  {\theta_{\rm pf}^2}\right)
\end{eqnarray}
is given by
\begin{eqnarray}
  q_{\rm pf}&=&
  -1-\frac{3}{\left[\frac{vv'}{1-v^2}+2\bqz+\bqp+3v\bc\right]^2}\times 
  \nonumber \\ &&
  \left(2\bqz' +\bqp'+\frac{v(2\bqz+\bqp)+\bc(3-v^2)}{1-v^2}v'+
  \frac{1+2v^2}{(1-v^2)^2}{v'}^2
  \right. \nonumber \\ &&
  +\frac{vv''}{1-v^2}
  -(2\bqz+\bqp+v')\left\{\bqp+\bqz\left[2(\bqp^2+\bu^2)
  \right. \right. \nonumber \\ && \left. \left.
  +\frac{\gamma}  {1+(\gamma-1)v^2}\Omega_n\right]\right\} 
  -\left.\left\{2(\bqz-\bqp)+3v\bc\right\}v\bc\right) .
\end{eqnarray}

\section{Transformation between the SSS and TSS variables}

The two sets of variables basically differ only in the choice of
dominant quantities (although one has to be careful, as the
change of causality may result in sign change). Defining
\begin{eqnarray}
  \zeta&=&\frac{\ty}{\by} \\
  &=&\sqrt{\bqz^2-\bw^2} \\
  &=&\frac{1}{\sqrt{\tqz^2-\tct^2}} ,
\end{eqnarray}
the relations between the variable sets become
\begin{eqnarray}
  &&
  \bqz=\zeta\tqz , \quad
  \bqp=\zeta\tqp , \quad
  \bc=\zeta\tc , \quad
  v=u^{-1} , \quad
  U=\zeta\tu , \\
  && \bw=\zeta\sqrt{\tqz^2-1} , \quad
  \tct=\zeta^{-1}\sqrt{\bqz^2-1} .
\end{eqnarray}

\section*{Acknowledgements}

AC would like to acknowledge financial support from NSERC of Canada.
MG would like to thank the Department of Mathematics and Statistics
at Dalhousie University for hospitality while this work was carried out.

\end{document}